\documentclass[11pt,a4paper]{amsart}
\usepackage{geometry}                
\usepackage{graphicx}
\usepackage{amssymb}
\usepackage{epstopdf}
\usepackage{fullpage}
\usepackage{color}
\usepackage{float}
\usepackage{caption}
\usepackage[table]{xcolor}
\usepackage{tablefootnote}

\usepackage{url}
\usepackage{morefloats}

\newcommand{\Gam}{\mathsf{G}}

\newcommand{\negbin}{\mathsf{NegBin}}

\newcommand{\Poi}{\mathsf{Poi}}

\definecolor{mattgreen}{RGB}{216,228,188}
\definecolor{mattred}{RGB}{218,150,148}
\definecolor{mattgrey}{RGB}{64,64,64}

\usepackage{graphicx} 

\usepackage{array} 
\usepackage{paralist} 

\begin{document}

\title{Spatiotemporal Detection of Unusual Human Population Behavior Using Mobile Phone Data}

\author{Adrian Dobra}
\address{Department of Statistics, Department of Biobehavioral Nursing and Health Systems, Center for Statistics and the Social Sciences and Center for Studies in Demography and Ecology, University of Washington, Box 354322, Seattle, WA 98195}
\email{adobra@uw.edu}

\author{Nathalie E. Williams}
\address{Department of Sociology and Jackson School of International Studies, University of Washington, Box 353340,Seattle, WA 98195}
\email{natw@uw.edu}

\author{Nathan Eagle}
\address{Department of Epidemiology, Harvard University, Boston, MA 02115}
\email{nathan@mit.edu}

\begin{abstract}
With the aim to contribute to humanitarian response to disasters and violent events, scientists have proposed the development of analytical tools that could identify emergency events in real-time, using mobile phone data. The assumption is that dramatic and discrete changes in behavior, measured with mobile phone data, will indicate extreme events. In this study, we propose an efficient system for spatiotemporal detection of behavioral anomalies from mobile phone data and compare sites with behavioral anomalies to an extensive database of emergency and non-emergency events in Rwanda. Our methodology successfully captures anomalous behavioral patterns associated with a broad range of events, from religious and official holidays to earthquakes, floods, violence against civilians and protests. Our results suggest that human behavioral responses to extreme events are complex and multi-dimensional, including extreme increases and decreases in both calling and movement behaviors. We also find significant temporal and spatial variance in responses to extreme events. Our behavioral anomaly detection system and extensive discussion of results are a significant contribution to the long-term project of creating an effective real-time event detection system with mobile phone data and we discuss the implications of our findings for future research to this end.
\end{abstract}

\maketitle

\tableofcontents

\section{Introduction}

Discrete emergency events, such as terrorist attacks and natural disasters, occur frequently around the globe and regularly cause massive destruction.  However, it is often the aftermath of these events, the disaster period, when the greater problems arise, given social, economic or political inabilities to cope with the event \cite{blaikie-et-2004}.  We often find large evacuation or migration streams, organized criminal or militia reprisals, spread of infectious diseases, and changes in local mobility and economic behavior \cite{boyle-2010,myers-et-2008,bissell-1983,shears-1991,morrow-et-1991,verpoorten-2009}.  These emergency events can occur anytime, anywhere, and often without warning.  Occurrences in rural areas or countries with poor communication and transportation infrastructure make it difficult to identify and respond to such emergencies in a timely and appropriate manner to avert full scale disaster.  Indeed, it can be days before accurate information about an event even reaches government or non-governmental organizations \cite{chua-et-al-2007}.  Delays in response can exacerbate the magnitude and length of the disaster period after an emergency event, resulting in serious epidemiological problems \cite{watson-et-al-2007}.

With the ultimate aim to decrease the humanitarian toll of post-event disasters, scientists have recently begun to understand that several relatively new sources of organically collected data, such as cell phone records, internet blogs, and Twitter, could provide real time or very quick identification of emergency events \cite{schoenharl-et-2006,candia-et-2008,miruna-butts-2008,akoglu-faloutsos-2010,kapoor-et-2010,bagrow-et-2011,bengtsson-et-2011,gething-et-2011,traag-et-2011,lu-et-al-2012,sutton-et-2013,gao-et-2014,aleissa-et-2014,young-et-2014}. Human behaviors such as mobility, migration, frequency of connection, and size of social networks can be estimated with these data. Dramatic changes in regular patterns of these behaviors could signal a response to an emergency event, and thus be used to identify when and even where an event has happened.

While these data are continuously collected by service providers and could ostensibly be made available, the tools for using such data for real-time event identification are still under construction. The broad purpose of this article is to contribute to the long-term goal of development of analytical tools for using mobile phone data to identify emergency events in real time. This can ultimately contribute to quicker humanitarian response and decreases in the severity of disasters. Specifically, we create a system for identifying anomalies in human behavior as manifested in mobile phone data, and discuss the correspondence between these anomalies and actual emergency and non-emergency events that might have caused them. 

Previous research has demonstrated that such analytical tools might be possible, by showing that natural and man-made emergency events, such as earthquakes or bombings, can be ÒseenÓ in dramatic increases in calling and mobility behaviors \cite{schoenharl-et-2006,candia-et-2008,akoglu-faloutsos-2010,kapoor-et-2010,bagrow-et-2011,bengtsson-et-2011,gething-et-2011,traag-et-2011,lu-et-al-2012,gao-et-2014,aleissa-et-2014,young-et-2014}.  These studies are able to show such anomalies by comparing behaviors to events where the character, time, and place of these events are already known. We build on these studies, but develop a blind system that is closer in nature to an actual event detection system. Instead of starting with the time and location of an event, then looking for anomalous calling behavior, we develop a behavioral anomaly detection system that identifies days with unusual calling or mobility behavior, as well as the location and geographic extent of these disruptions. Our detection system is scalable as it is able to efficiently process years of country-wide mobile phone records.

For illustration we use mobile phone records from a single cellular services provider from Rwanda. We connect the identified anomalous days and locations with extensive records of violent and political events and natural disasters. Results of this exercise reveal that some days with anomalous increases in calling and mobility behavior match well with several different kinds of events. In other cases, days with {\it decreases} in calling and/or mobility match with events. These cases were surprisingly more numerous than events matched with increases in calling and mobility. In still other cases, we do not find good event matches for days with anomalous behavior and we also find cases where emergency events occurred without resulting in anomalous behavior that our system could detect. Notably, we learn as much from the unmatched events and behavioral anomalies as from the matched cases.  

We argue that further quantitative and qualitative research into the exact and possibly multi-dimensional nature of human response to emergency events is needed. In this regard, our careful analysis of both the matched events and the events and instances of anomalous behavior that do not match reveal some key insights into further developments needed to better understand human response to emergency events. In fact, it is this outcome, namely the demonstration that human behavioral responses to emergency events are much more complex than previously assumed, that is the most important contribution of this paper.  Future research must address this complexity and can benefit from using existing social and psychological theories of behavioral response to threat.    We conclude this article by setting out a clear pathway of research aimed at the goal of a creating an effective system of identifying emergency events in real-time (or close to real time) from mobile phone data.

\section{Materials and Methods}

\subsection{Measuring human behavior with mobile phone data}

Cellular service providers continuously collect mobile phone records for billing purposes and to improve the operation of their networks \cite{gonzales-et-al-2008,song-et-al-2010,becker-et-al-2013}. Every time a person makes a voice call, sends a text message or goes online from their mobile phone, a call detail record (CDR) is generated which records time and day, duration and type of communication, and an identifier of the cellular tower that handled the request. We analyze anonymized CDRs provided by a major cellular phone service provider in Rwanda. These data comprise all mobile phone activity in the providerÕs network between June 1, 2005 and January 1, 2009 \cite{blumenstock-2012,blumenstock-eagle-2012}.

Many of the existing methods for emergency event detection rely on call volume, at either the cellular tower level or at individual level \cite{candia-et-2008,kapoor-et-2010,bagrow-et-2011,gao-et-2014,aleissa-et-2014,young-et-2014}, as the sole measure of human behavioral response. However, the number of calls is only one type of behavior that could change in response to emergency events.  Several studies have demonstrated that population mobility is also severely affected by large-scale disasters \cite{bengtsson-et-2011,lu-et-al-2012,wolley-et-2013}, thus mobility should also be considered to improve the efficiency and reach of event detection systems.  For example, some events, such as tsunamis, might require immediate evacuation and leave time to make phone calls only after the event is over. In this case, we might find initially increased mobility but decreased call frequency.  

From the Rwandan mobile phone data, we create two measures of behavior: call frequency and movement frequency.  For both measures, we chose a day as the reference unit of time, so our measures are the number of calls per day and number of moves per day. Our data provide 327,335,422 person days of each measure. Periods of time that are shorter or longer than a day can be employed without any subsequent changes to our methods.

Call frequency is a relatively straightforward measure, whereas measuring movement frequency is more involved, given the complexities of defining what is a ``move" using mobile phone data. First, a person's path of travel for a whole day must be traced; we call this trace a spatiotemporal trajectory.  The approximate spatiotemporal trajectory of a mobile phone and its user can be reconstructed by linking the CDRs associated with that phone with the locations (latitude and longitude) of the cellular towers that handled the communications. Instead of defining spatiotemporal trajectories directly with respect to the locations of the cellular towers, we use a system of 2040 grid cells each measuring 5 km x 5 km that covers Rwanda's territory \cite{williams-et-2014}. Some grid cells have a cellular tower in them, some do not, and some have multiple cellular towers. We refer to a grid cell with at least one active tower as a {\it site}. The introduction of a grid system increases error in location measurement slightly, but is necessary to alleviate serious problems of endogeneity between mobility measurements and social, economic, and political characteristics of context and spatial placement of mobile phone towers. Consistent use of 5 km x 5 km cells, instead of cells of other sizes, is also necessary so as not to create problems similar to the modifiable areal unit problem (MAUP) \cite{fotheringham-wong-1991}.  See \cite{williams-et-2014} for a detailed discussion on these issues. Once a grid system is imposed and a spatiotemporal trajectory created for each person, movement frequency can be calculated as the number of times a person makes a call from a different grid cell than the previous call \--- see SI Appendix, Section SI1 for details. 

\subsection{Event records}

Our data on violent and political events, natural disasters, and major holidays come from a variety of public sources. We use an existing dataset of violent and political events from the Armed Conflict Location and Event Data Project (ACLED)\footnote{Accessible at \url{http://www.acleddata.com/}}. ACLED collects extensive data on conflict-related events including battles, killings, riots and protests, and violence against civilians. Their information, obtained from local and international newspaper and radio sources, includes details on the date and location of each event, as well as the type of event, groups involved, and fatalities. We use ACLED data from Rwanda and provinces that border Rwanda in Burundi, Democratic Republic of Congo (DRC), Uganda, and Tanzania. Data on natural disasters come from Reliefweb\footnote{Accessible at \url{http://reliefweb.int/}} which provides the location, date, extent of damage, and further details of a variety of natural disasters around the world, from storms, to volcano eruptions, floods, heatwaves, insect infestations, and earthquakes. We supplement these data sources by searching on the internet and in Rwandan newspapers (e.g., New Times of Rwanda\footnote{Accessible at \url{http://www.newtimes.co.rw/}} and Rwanda Focus\footnote{Accessible at \url{http://focus.rw/wp/}}) for events that might explain what happened during the days on which we find anomalous calling and mobility behaviors. 

\begin{figure}[htbp!]
\begin{center}
\includegraphics[width=5.5in,angle=0]{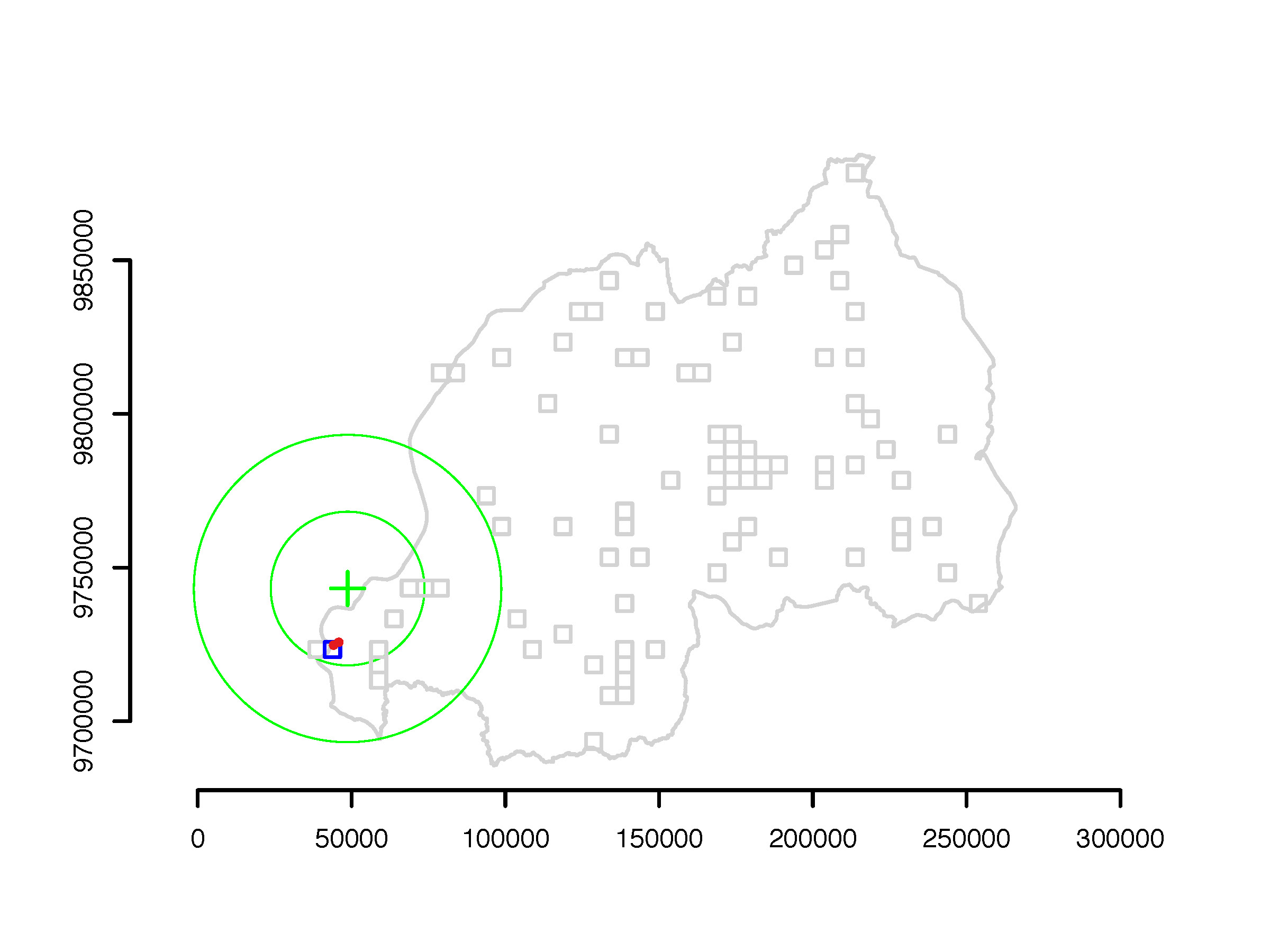}
\caption{{\bf Location of the site with index 361.} The three cellular towers (red dots) that were active in January and February 2008 and were located in site 361 recorded higher than usual call volume and movement frequency on February 3, 2008 \--- the day of the Lake Kivu earthquakes. The green cross marks the approximate location of the epicenters of the Lake Kivu earthquakes, and the two green circles mark the 25 and 50 km areas around the epicenters. The location of site 361 is shown in blue, while the locations of the other 84 sites that contained active towers in February 2008 are shown in gray. The Rwandan country borders are also shown in gray.}
\label{fig:sitesEarthquake}
\end{center}
\end{figure}

Amongst all events in our dataset, to exemplify our anomalous behavior detection system, we use a series of large earthquakes whose epicenters were located in the south part of Lake Kivu region. The earthquakes occurred between 9:34 am and 1:05 pm local time on Sunday, February 3, 2008 and struck parts of Rwanda, the Democratic Republic of Congo (DRC), and Burundi, leaving 44 people dead and hundreds injured. Figures \ref{fig:sitesEarthquake}, \ref{fig:UPFebruary32008} and \ref{fig:DOWNFebruary32008} show the location of the epicenters of the earthquakes as well as the locations of the sites with active cellular towers in that time period. In particular, site 361 (Figure \ref{fig:sitesEarthquake}) is one of the closest sites to the epicenters of the Lake Kivu earthquakes and contains three cellular towers. The active sites located in a 50 km radius from the epicenters contain at most two towers.

\begin{figure}
\begin{center}
\includegraphics[width=5.5in,angle=0]{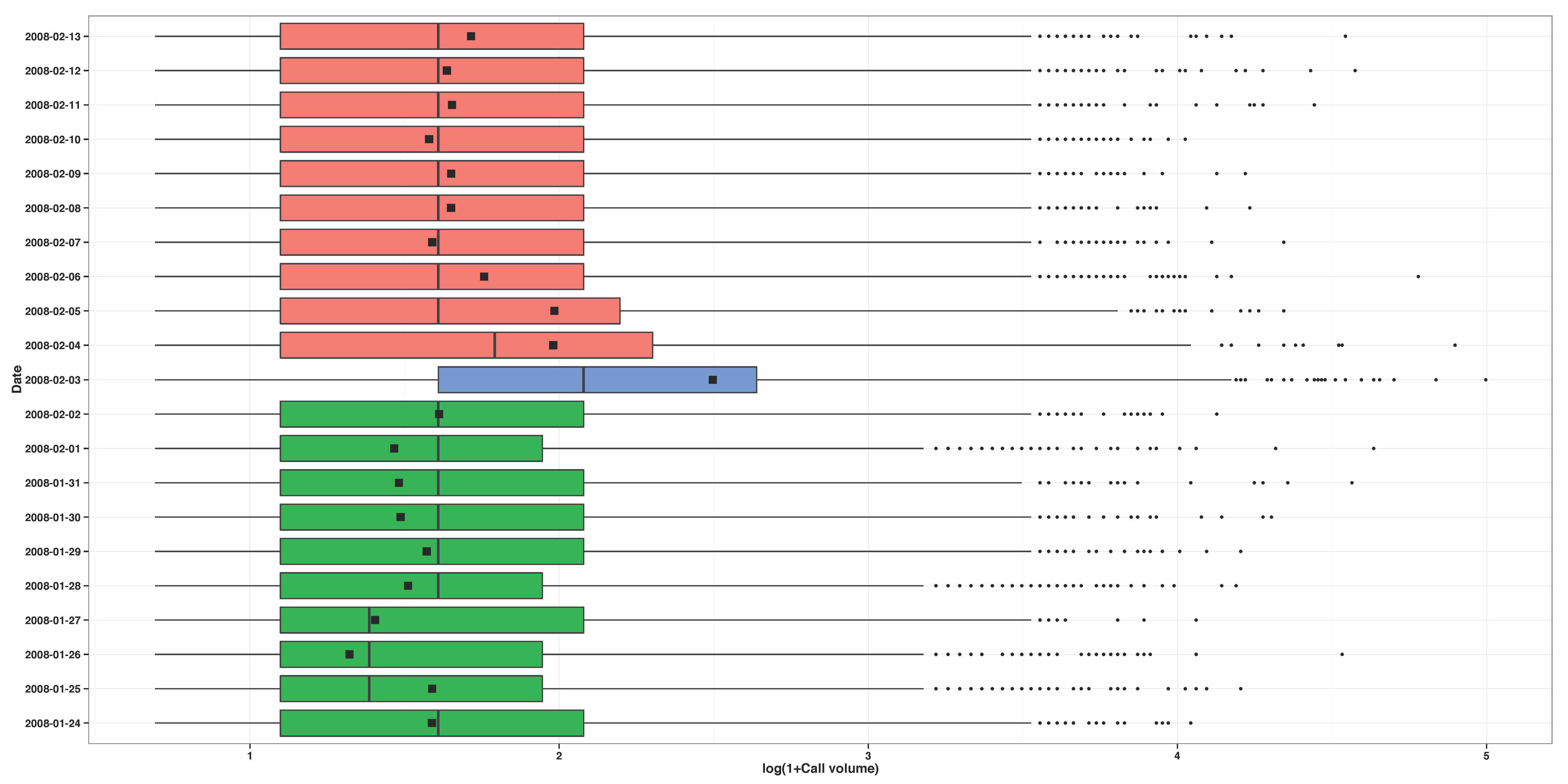}
\caption{
{\bf Call volume for site 361.} Calling behavior of the people who made at least one call from at least one cellular tower located in site 361 between January 24, 2008 (10 days before the Lake Kivu earthquakes) and February 13, 2008 (10 days after the Lake Kivu earthquakes). The side-by-side boxplots represent the distribution of the number of calls made by these people in each of the 21 days. The squares indicate the total number of calls made in site 361 in each of the 21 days.
}
\label{fig:lakeKivuSite361Callvolume}
\end{center}
\end{figure}

Our system is based on call and movement frequency in a particular place; thus our unit of analysis is a site, not a person. As such, once call frequency and movement frequency are calculated for each person, we associate the daily spatiotemporal trajectory of each caller with every site from which they made at least one call that day. Figures \ref{fig:lakeKivuSite361Callvolume} and \ref{fig:lakeKivuPlace361Numberoftrips} show the distributions of the call and movement frequency measures of the callers that placed at least one call from one of the three cellular towers located in site 361, 10 days before and 10 days after the day the Lake Kivu earthquakes occurred. It is clear that the earthquakes had a significant impact in the lives of the people who made calls from site 361: during the day of the earthquakes, users of the towers from this site made more calls and were more mobile compared to users of the towers from the previous 10 days and the next 10 days.

\subsection{Understanding emergency events and behavioral response possibilities}

Existing approaches for identifying abnormal patterns of human behavior from mobile phone data focus almost exclusively on the following scenario \cite{candia-et-2008,akoglu-faloutsos-2010,kapoor-et-2010,bagrow-et-2011,gao-et-2014,aleissa-et-2014,young-et-2014}: a group of people $G_0$ happen to be close to the location of an emergency event $\mathcal{E}$ and, as a result of their witnessing the event, start communicating with their family and friends about the event. Another group of people $G_1$ directly reached by the members of $G_0$ could, in turn, further communicate about $\mathcal{E}$ with people in $G_0\cup G_1$ or with other people. Information about $\mathcal{E}$ propagates through contact networks and media outlets to reach even more people. These outgoing and incoming communications (calls, text messages, social media posts) trigger a spike in the mobile phone activity of the members of $G_0$ immediately following $\mathcal{E}$. Since group $G_0$ is assumed to be spatially close to $\mathcal{E}$, the methods presented in \cite{bagrow-et-2011,gao-et-2014,aleissa-et-2014,young-et-2014} proceed by assuming that the time, duration and location of several emergency events are known. Based on the exact spatiotemporal localization of $\mathcal{E}$, they identify the cellular towers $\mathcal{T}$ in the immediate proximity of $\mathcal{E}$, and find out the corresponding groups $G_0$ of people that made calls from these towers in a time frame which spans the time of occurrence of $\mathcal{E}$. They subsequently analyze the time series of total outgoing and incoming call volumes at towers in $\mathcal{T}$ to show that, as expected, there is an increased number of calls immediately following $\mathcal{E}$ and present statistical models that are able to identify the communication spikes.

However, when creating a system to blindly identify emergency events without prior knowledge that they occurred, more understanding of events and behavioral response possibilities is required. We discuss a few here, but there are many other dimensions that will likely be discovered as the literature on behavioral response to emergency events grows. First, when looking for anomalous behaviors, we must address routine behavioral patterns.  For example, people routinely make more and fewer phone calls and are more and less mobile during particular times of day and night and on different days of the week and month \cite{ratti-et-2006,liu-et-2009}.  Mobile phone systems also progressively service more users and build more towers over time. In Rwanda, for example, the changes in the mobile phone system over time are non-linear, and sometimes dramatic \cite{williams-et-2014}. Consequently, anomalous behaviors would not be just dramatic changes over time in calling or mobility, but would be changes compared to routine behaviors after the temporal variance in numbers of users and towers is taken into account. 

\begin{figure}
\begin{center}
\includegraphics[width=5.5in,angle=0]{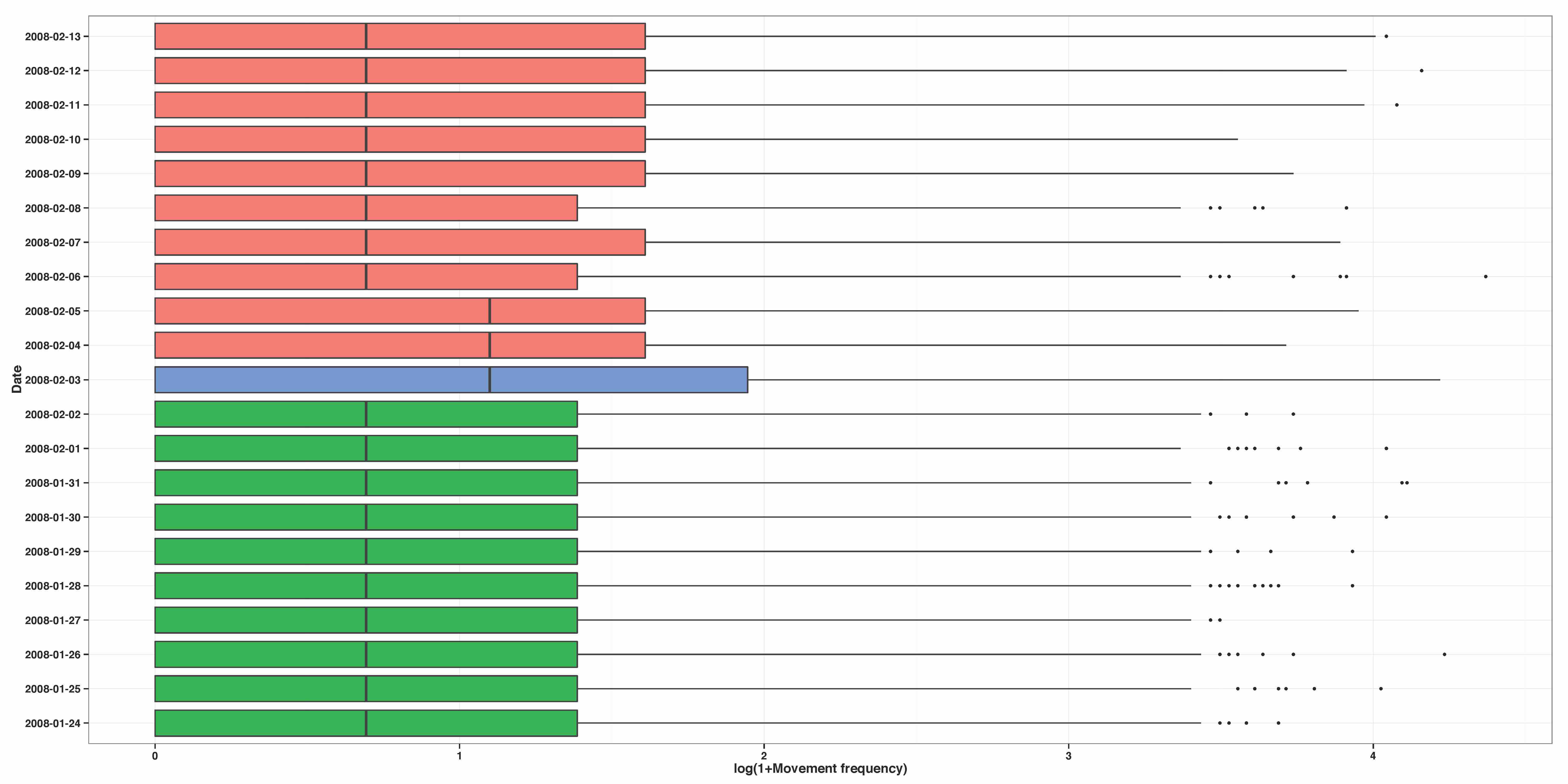}
\caption{
{\bf Movement frequency for site 361.} Mobility behavior of the people who made at least one call from at least one cellular towers located in site 361 between January 24, 2008 (10 days before the Lake Kivu earthquakes) and February 13, 2008 (10 days after the Lake Kivu earthquakes). The side-by-side boxplots represent the distribution of the movement frequency of these people on each of the 21 days.}
\label{fig:lakeKivuPlace361Numberoftrips}
\end{center}
\end{figure}

The situation is further complicated by the reality that many emergency events, however discrete in time, are followed by longer periods of disaster, characterized by breakdowns in social, political, and economic systems \cite{blaikie-et-2004}. This creates a situation where new routine behaviors in the post-event disaster period might be quite different from routine behaviors in a pre-event period. This is shown in Figure \ref{fig:lakeKivuPlace361Numberoftrips}, with less stable mobility after the Lake Kivu earthquakes compared to before.  Thus it is the brief period of time during and just after an emergency event when we expect to find the largest changes in reactionary behaviors, and it is the longer pre-event and post-event disaster periods to which we must compare.

Second, in addition to emergency events, planned non-emergency events occur often and these can disrupt routine behavioral patterns as well. \cite{bagrow-et-2011} find dramatic changes in call frequency in response to festivals and concerts, and it is likely that other events, including holidays, will also produce changes. The period of time in which we can expect to find the largest change in reactionary response to a planned event (in contrast to unplanned events) could include the immediate pre-event period, the event itself, and the immediate post-event period. If our goal is to identify emergency events using changes in behavioral patterns, we must also identify, and separate, non-emergency events that could also produce behavioral changes.

Third, it is possible that there is more than one emergency or non-emergency event in a single day.  Different events could influence people in a small area, in a region, or even across a whole country.  An effective event identification system must be able to identify when behavioral patterns suggest a single localized event, multiple localized events, or a single event that produces behavioral responses over a wide area. 

Fourth, behavioral responses to an emergency event could include dramatic increases but also dramatic decreases in call frequency or mobility behavior. Broad assumptions, backed up by some evidence in the literature \cite{bagrow-et-2011,lu-et-al-2012}, suggest that people will call more and become more mobile during and after emergency events. The logic underlying this belief is that people will call others to tell them about the event and will move away from any danger. This might be true for some cases and on some time scales. An alternate possibility is that initial evacuation or escape from a dangerous situation, such as a tsunami, could preclude the ability to make a phone call. In this case, we would find increased mobility but decreased call frequency. Another possibility is a situation where an emergency event, such as a flash flood destroys roads or other transportation infrastructure, forcing people to stay in place and disrupting other daily routines. In this second case, we might find decreased mobility but increased call frequency. Thus, there are strong theoretical reasons to expect dramatic decreases in certain behaviors in the immediate aftermath of some emergency events. An effective event detection system must identify both increases and decreases in both call and movement frequency.

\begin{figure}[htbp!]
\begin{center}
\includegraphics[width=5in,angle=0]{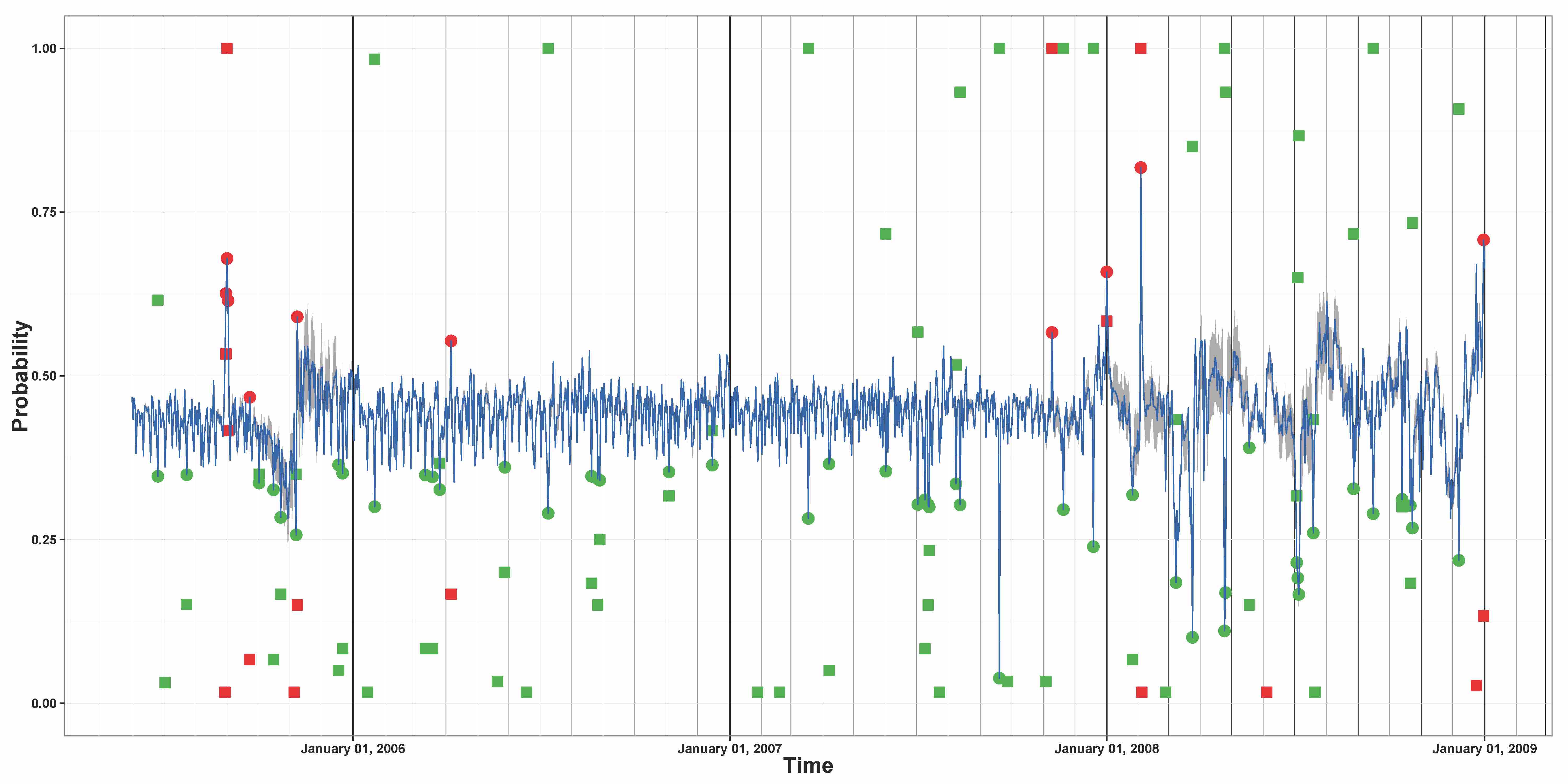}
\caption{{\bf Daily call volume time series of probabilities associated with site 361.} The means of the estimated probabilities of making more calls are shown in blue. The gray band gives the intervals between the 2.5\% and the 97.5\% quantiles of the estimated probabilities. The dots indicate which days in this time series are extreme positive outliers (red) and extreme negative outliers (green). The red (green) squares indicate the confidence probabilities that a day is a positive (negative) extreme outlier. The confidence probabilities are shown only for the days that have been classified as an outlier at least once.}
\label{fig:Site361CallVolume}
\end{center}
\end{figure}

\begin{figure}[htbp!]
\begin{center}
\includegraphics[width=5in,angle=0]{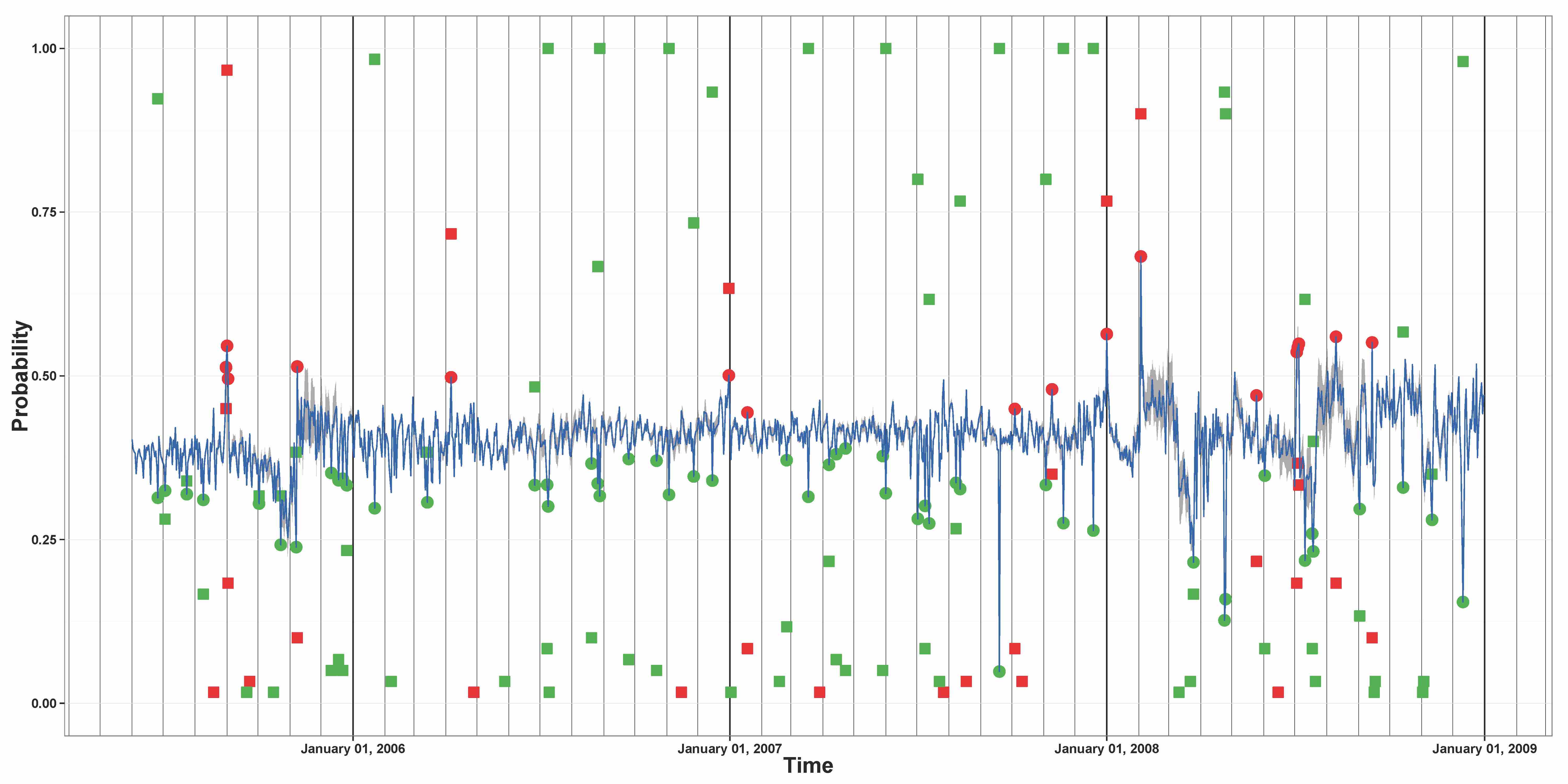}
\caption{{\bf Daily movement frequency time series of probabilities associated with site 361.} The means of the estimated probabilities of being more mobile are shown in blue. The gray band gives the intervals between the 2.5\% and the 97.5\% quantiles of the estimated probabilities. The dots indicate which days in this time series are extreme positive outliers (red) and extreme negative outliers (green). The red (green) squares indicate the confidence probabilities that a day is a positive (negative) extreme outlier. The confidence probabilities are shown only for the days that have been classified as an outlier at least once.}
\label{fig:Site361MovementFrequency}
\end{center}
\end{figure}

\subsection{Identifying behavioral anomalies}

Our proposed approach for detecting abnormal communication patterns is designed to capture not only days and regions with higher than usual call frequency, movement frequency, or both, but also days and regions with lower than usual levels of these behaviors. The assessment is performed longitudinally at the site level in the first step of our system. At the second step, the disruptions from the first step are combined across sites for each day, allowing us to determine the spatial extent of behavioral anomalies and if there was more than one possible event on the same day.\\

{\bf Step 1: Identifying days with anomalous human behavior at one site}. In order to separate anomalous from routine behaviors (both pre-event routine or post-event routine behaviors), we create reference periods of time. We divide the set of days with cellular communication data from a site into subsets of T consecutive days. The length T of the reference time periods is important and must be selected based on two considerations: (i) T should be sufficiently small such that fluctuations in the number of active towers and the number of callers in each site during T consecutive days are not excessive; and (ii) T should be sufficiently large such that the effects of emergency events and the post-event disaster period are reasonably low with respect to periods of T consecutive days. After a close examination of the temporal dynamics of the cellular network of the provider of the Rwandan CDRs and of the types of events that we know to have occurred between June 1, 2005 and January 1, 2009, we decided to use T=60. 

We consider each period P of T consecutive days with available CDRs from a site S. For each day t in P, we look at the spatiotemporal trajectories of callers who made at least one call from S. We use Poisson models to estimate: (1) the probability that a random caller on day t made more calls than a random caller on a random day in P other than t; and (2) the probability that a random caller on day t moved more frequently than a random caller on a random day in P other than t. Our estimation method of these two probabilities is detailed in SI Appendix, Section SI3. An event that increases (decreases) the call volume or mobility of callers during day t is associated with unusually high (low) probabilities of making ore calls or moving more frequently. To identify such days in the call volume and movement frequency time series of estimated probabilities, we fit beta regression models \cite{cribari-neto-zeileis-2010} with time as the explanatory variable and the estimated probabilities as the response variable, and determine which days are positive or negative outliers based on standardized weighted residuals 2 \cite{espinheira-et-2008}. Estimates of probabilities of making more calls and of moving more frequently are produced for a day t and a site S with respect to each reference time period of length T that day t belongs to. The behavior of callers during day t at site S could be classified as unusual with respect to a reference time period, or as normal with respect to another reference time period.

We define the confidence probability that call or movement frequency are unusually high or low on day t as the ratio between number of times the corresponding probability estimates have been classified as positive or negative outliers and the number of reference time periods used to produce these estimates. Any day with confidence probability less than a threshold, we use 0.05, is classified as an extreme outlier day. Figures \ref{fig:Site361CallVolume} and \ref{fig:Site361MovementFrequency} show the time series of the two types of daily probabilities for site 361. The figures present the confidence probabilities for those days that were classified as positive or negative outliers at least once. The extreme positive and negative outliers are also shown. February 3, 2008 \--- the day of the Lake Kivu earthquakes \--- is among the extreme positive outliers for both the call volume and the movement frequency measures for site 361. We note that there are more extreme negative outliers than extreme positive outliers which means that there are more days in which the call volume or movement frequency at site 361 was unusually low than days in which the call volume or movement frequency at site 361 was unusually high. In fact, a similar pattern is present in call volume and movement frequency time series associated with most of the other Rwandan sites.

The output from Step 1 of our approach is a set of two time series (one for call frequency and one for movement frequency) that cover the entire study period, for each site in the study area. In our study, there were 155 sites that were active at some time during the study period, thus our output was 310 time series, together with their corresponding sets of extreme positive and negative outlier days. These are days when anomalous behavior occurred, at each site separately. This output provides no information about the spatial extent of behavioral anomalies (whether the anomaly occurred at one site or many) and the likelihood that anomalies at different sites were related or not. For this information, we continue to Step 2 of our method.\\

\begin{figure}
\begin{center}
\includegraphics[width=5.5in,angle=0]{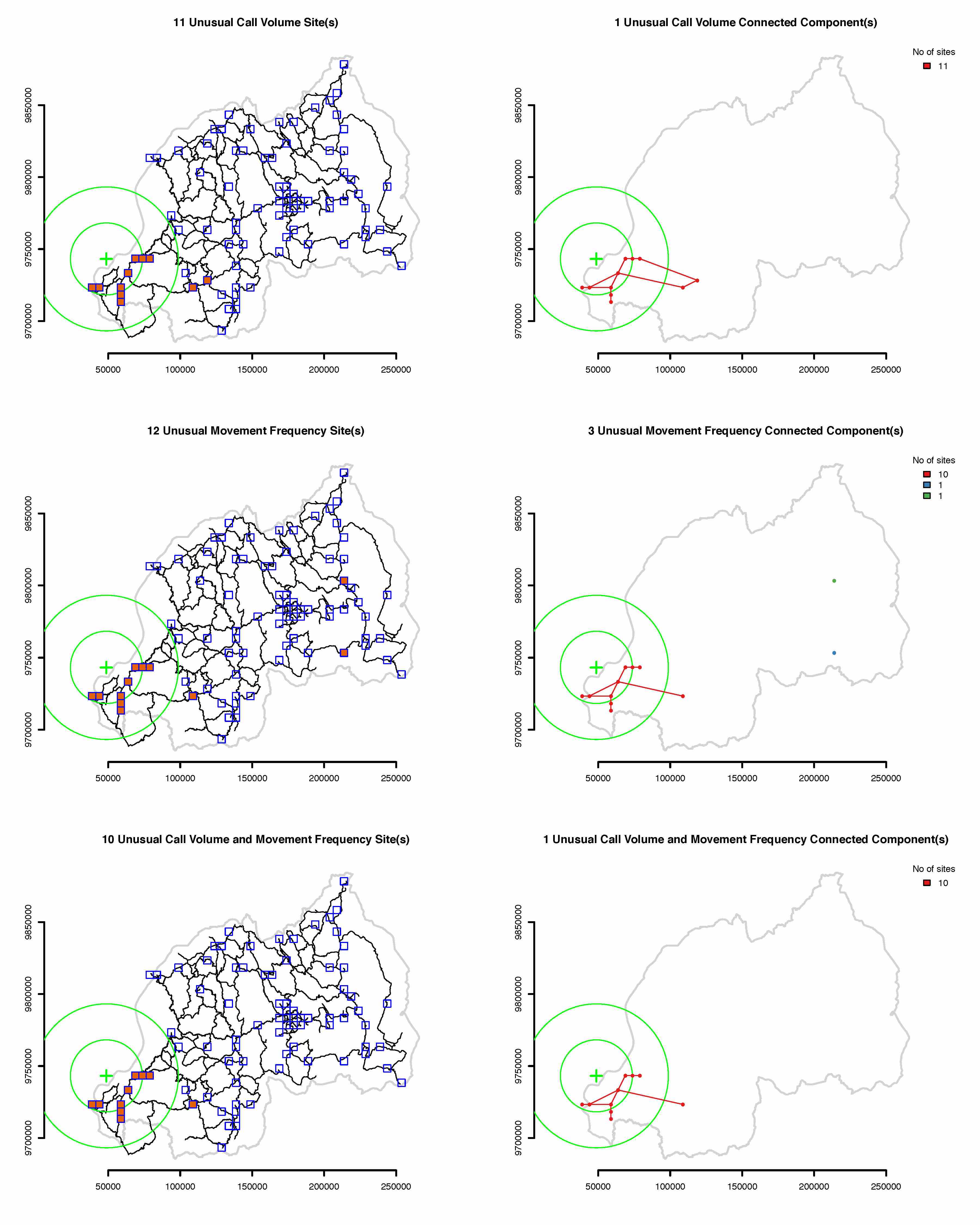}
\caption{
{\bf Sites with unusually high behavior on February 3, 2008.}  The green cross marks the location of the epicenters of the Lake Kivu earthquakes, while the two green circles mark the 25 and 50 km areas around the epicenters. Ten sites recorded unusually high call volume and movement frequency and belong to the same spatial cluster. One additional site recorded unusually high call volume, while two additional sites recorded unusually high movement frequency. Most of these sites are located within 50 km of the approximate location of the earthquakes epicenters which is indicative of the connection between the anomalous pattern of communication and the occurrence of the Lake Kivu earthquakes.
}
\label{fig:UPFebruary32008}
\end{center}
\end{figure}

\begin{figure}[htbp!]
\begin{center}
\includegraphics[width=5.5in,angle=0]{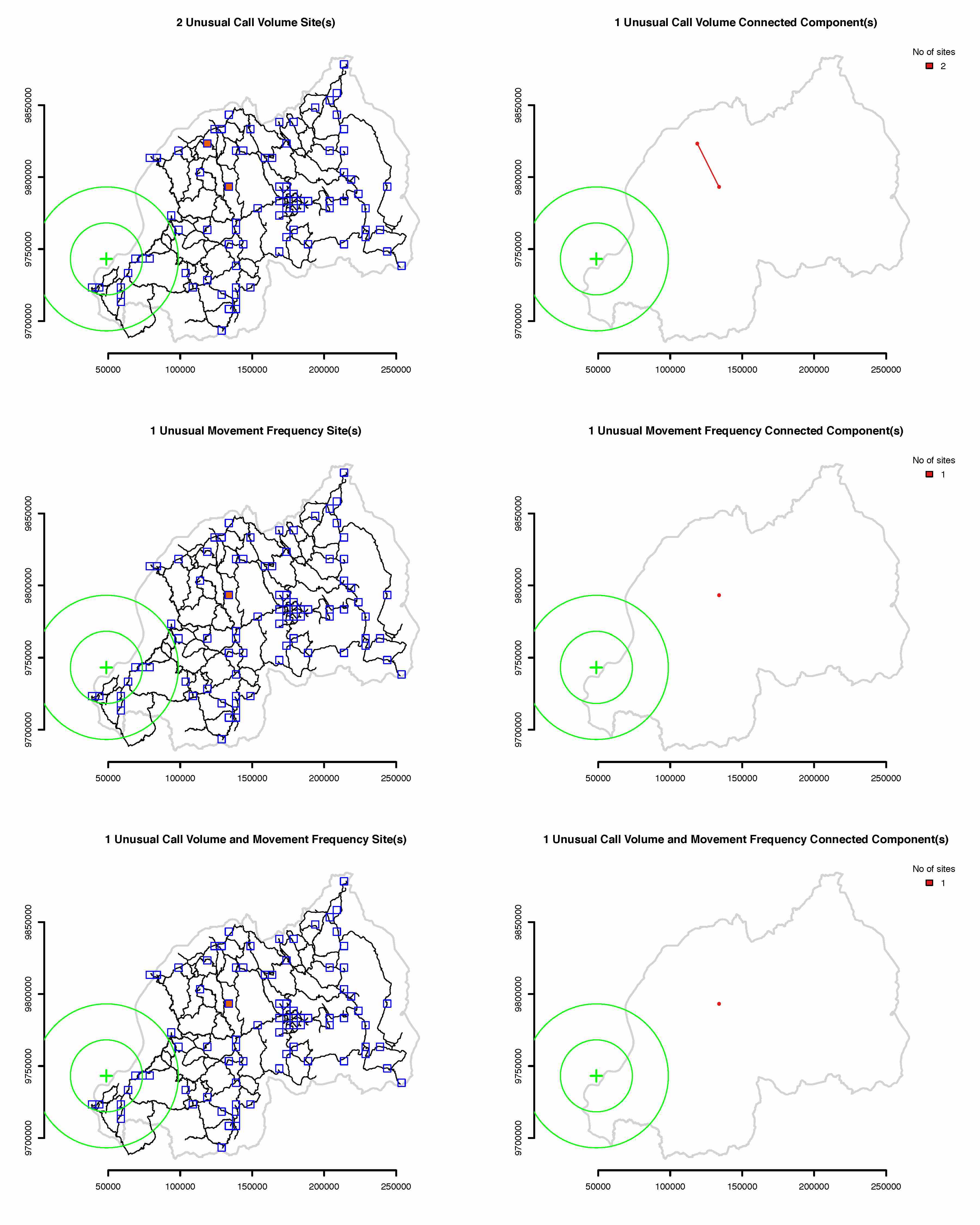}
\caption{{\bf Sites with unusually low behavior on February 3, 2008.}  The green cross marks the location of the epicenters of the Lake Kivu earthquakes, while the two green circles mark the 25 and 50 km areas around the epicenters. One site recorded unusually high call volume and movement frequency, and one additional site recorded unusually high call volume. Both sites belong to the same spatial cluster, and are located relatively far from the approximate locations of the earthquakes epicenters. The anomalous pattern of communications at these two sites could be caused by some other event, possibly unrelated with the Lake Kivu earthquakes.
}
\label{fig:DOWNFebruary32008}
\end{center}
\end{figure}

{\bf Step 2: Identifying days with anomalous human behavior at multiple sites}. For the second step of our approach, we create maps that display, for every day, the sites for which that day is an extreme positive or negative outlier. Figures \ref{fig:UPFebruary32008} and \ref{fig:DOWNFebruary32008} present these maps for February 3, 2008. We construct and discuss similar maps for other days with extreme outliers in the Results section. 

Our maps are designed to facilitate the identification of spatial clusters of sites that are extreme outliers of the same type in the same day. For each day there are two series of maps, one for extreme positive outliers, another for extreme negative outliers. Each has six panels grouped in three rows (sites with unusual call frequency, sites with unusual movement frequency and sites with both). The maps in the first column show the locations of the sites with disturbances and the locations of the sites without disturbances. Sites that are extreme outliers of the same type, positive or negative, and are on average spatially closer to each other than to other sites, comprise a cluster. We suggest that the anomalous behavior in a spatial cluster of sites is most likely caused by a single event. Separate spatial clusters are more likely to represent different events in different places.

The two maps in the third row of Figure \ref{fig:UPFebruary32008} show 10 sites with unusually high call volume and movement frequency. Nine of these sites are all the sites located in Rwanda within a 50 km radius from epicenters of the Lake Kivu earthquakes, thus it is very likely that this unusual behavioral pattern was caused by the earthquakes. The maps in the first row of Figure \ref{fig:DOWNFebruary32008} show two sites with unusually low call volume, and the maps on the second and third row show that one of these sites also recorded unusually low movement frequency. Despite these unusual behavioral patterns occurring on the same day as the Lake Kivu earthquakes, they were likely caused by another event of a different type, not only because they led to lower rather than higher than usual call and movement frequency, but also because they occurred far from the epicenters of the earthquakes.

Instead of simple visual examination of the maps, we use a systematic method to identify spatial clusters of sites. Our goal is to place sites together in a cluster if they exhibited the same kind of unusual behavior (positive or negative) on the same day and if they are very close together or directly connected by a road with no other sites in between. To do this, we create a spatial neighborhood graph $\mathcal{G}$ as follows. Each site is associated with a vertex in $\mathcal{G}$. Two vertices are connected by an edge in $\mathcal{G}$ if the two corresponding sites are neighbors in the grid cell system or in the road network system we constructed for Rwanda \--- see SI Appendix, Section SI1. Two sites are neighbors in the grid cell system if they share an edge or a corner of the grid cells that define them. The road network system connects any two sites based the quickest road paths between their centroids, and does not contain loops, i.e., routes that leave one site, then return to the same site before reaching another destination site. We define two sites as neighbors in the road network system if there does not exist any other site on the quickest route path between them.

For a set of sites that had unusual behavior on a particular day, $\mathcal{A}$, the spatial neighborhood graph $\mathcal{G}$ induces a subgraph $\mathcal{G}(\mathcal{A})$. This is also a graph whose vertices are the set of sites $\mathcal{A}$ and whose edges connect those sites in $\mathcal{A}$ that are also linked in $\mathcal{G}$. As opposed to  $\mathcal{G}$, a subgraph $\mathcal{G}(\mathcal{A})$ is not necessarily connected: there could exist sites in $\mathcal{A}$ that are not linked by a sequence of edges in this subgraph. The connected components of $\mathcal{G}(\mathcal{A})$ (i.e., subgraphs of this graph that are connected) represent the spatial clusters of sites identified by our system. 

Consider the maps from the second row of Figure \ref{fig:UPFebruary32008}. There are 12 sites with higher than normal movement frequency, and 10 of these sites located in the proximity of the Lake Kivu earthquakes belong to one large spatial cluster. The disturbances at these sites were probably caused by the earthquakes. The two remaining sites located in western Rwanda are farther away from the other 10 sites and from each other, and belong to two other spatial clusters. The anomalous patterns of behavior at these two sites are likely to have been caused by events different than the Lake Kivu earthquakes.

The number and location of spatial clusters is important to estimate the number and location of outlier events. The number of sites in a single spatial cluster is a key indicator of the possible spatial reach of an event. The more sites in a cluster, the more wide-ranging were the behavioral anomalies, suggesting the wider was the range of the population that was influenced by an event. From another perspective, the largest spatial clusters of sites indicate the most significant events with the largest impact. For example, Figures \ref{fig:UPFebruary32008} and \ref{fig:DOWNFebruary32008} show spatial clusters with one, two, and eleven sites per cluster. The impact of the events that caused these clusters was reasonably concentrated in space. 

\section{Results}

Our anomalous behavior detection system identified many days with unusual calling and movement behavior across multiple sites. Figures S2-S8 in SI Appendix show daily time series of the size of the largest spatial clusters of sites that were extreme positive or negative outliers for one or both of our behavioral measures, call and movement frequency. There are numerous days in which the largest spatial clusters comprise 20, 30 or even more than 60 sites and cover much of the entire country. Here we describe some of these disturbances in which the largest spatial clusters of sites were identified, match some of them with key events that occurred in Rwanda, and discuss what we learn from each matched and unmatched event. This serves as a test of how well our system identifies key events in Rwanda and to highlight the next steps towards the goal of creating an effective emergency event identification system with mobile phone data. As discussed earlier, although our aim is to identify emergency events, non-emergency events might also cause behavioral changes. It is important to understand the particular behavioral signatures of many different events, in order to effectively identify and differentiate those that are emergencies. In this regard, our system identified several kinds of events, from hazards to holidays, all of which we now discuss. 

{\bf Violence against civilians \--- September 16, 2005} (Figure S9, SI Appendix). Our system identified a spatial cluster of four sites with higher than usual call volume and movement frequency on September 17, 2005. These sites are located in the vicinity of the city of Goma, along the Rwanda-DRC border and near the Rwanda-Uganda border. We find no event, emergency or otherwise that occurred on September 17, 2005. However, a violent event, reported by ACLED, occurred in the same area on September 16, 2005. The distance between the event's reported location (latitude -1.67, longitude 29.22) and the centroid of the closest site in the spatial cluster is 1.7 km. Radio France reported that this event involved five armed attacks in which 1 person was killed, 7 others were wounded, and was part of the general trend of violence against civilians. Upon close consideration, the one day lag between the event and the behavioral changes is reasonable. It is likely that the attacks were directly witnessed by a relatively small proportion of the local population who then communicated this information to others and the information propagated from there. As such, it could take a full day until a large proportion of the local population heard about and reacted to the event. 

From this case we gain a two key insights. First, our system can identify behavioral anomalies that are reasonable matches to emergency events. Second, information propagation might take time to reach a large portion of an affected population and thus behavioral anomalies that are identifiable at the population level might occur well after an event.

{\bf Violence against civilians \--- January 15, 2006} (Figure S10, SI Appendix). We identify 16 sites
(eight in the Kigali area) with lower than usual call volume on January 15, 2006. Eight of these sites also had lower than usual movement frequency. A violent event was recorded by ACLED in Kigali on the same day. The distance between the event's reported location (latitude -1.96, longitude 30.04) and the centroid of the closest site with unusual call volume is 3.8 km. Reporters Without Borders and Amnesty International report that four armed men invaded and ransacked the home of Bonaventure Bizumuremyi, the editor of the independent fortnightly Rwandan newspaper, Umuco. Mr. Bizumuremyi was the target of intimidation and harassment, demanding that he stop publishing articles criticizing the ruling Rwandan Patriotic Front (RPF). The armed forces of the RPF forced many independent journalists and human rights activists to leave Rwanda during that same general time period by intimidation, harassment or arbitrary arrest\cite{amnestyInternational-2006}. We find no records of other events on this day or those prior to it. It is possible that a large proportion of people in the Kigali area and in the other sites in the spatial cluster felt threatened or incensed by these actions against a public figure Mr. Bizumuremyi.

Note that the behavioral anomalies we find on this day are {\it decreases} in both call and movement frequency. Although it is easy to assume that people will flee violence, another possibility is that when threatened, people stay home and away from public areas. This explanation is similar to evidence from Nepal showing decreased migration following bomb blasts \cite{williams-et-2012}. Further in-depth research will be necessary to understand the exact connections between violence, threat, mobility, and calling behavior. In the present, we learn from this case that violent events might influence dramatic reductions in call frequency and mobility, a key contribution of this study to event detection and to our understanding of human response to violence and threat.

{\bf Protest \--- November 25, 2006} (Figure S11, SI Appendix). Our system identified seven sites with unusually low call volume and movement frequency, and seven additional sites with unusually low movement frequency on November 26, 2006. One of these sites is far from the other 13 and belongs to a separate spatial cluster. This suggests that there were two events on this day, one that created the anomaly in a single site and another that created behavioral anomalies in the remaining 13 sites. The 13 cluster sites are in Kigali and slightly to the east of the city. The single separate site is in the southern part of Rwanda on the Burundi border. We do not have record of any events that occurred on November 26, 2006 in these areas.  However, a large protest was recorded on November 25, 2006 in Kigali. The distance between the event's reported location (latitude -1.96, longitude 30.04) and the centroid of the closest site with unusual call volume (movement frequency) is 3.38 (1.83) km. The New Times of Rwanda reports that 15,000 demonstrators flooded the streets of RwandaÕs capital Kigali in protest of France's role in the Rwandan genocide, and their call for the arrest and trial of the Rwandan President Paul Kagame. Again, the response we find is decreased call and movement frequency and again it is the day after a significant event. It is possible that the large demonstrations created an atmosphere of threat and uncertainty and concerns about reprisal. This could lead people to stay at home and away from public areas, disrupt daily routines, and generally lead to less mobility.  Studies on fear of violence find similar patterns of behavior, where residents of unsafe neighborhoods spend more time indoors and away from public areas where violence could occur \cite{whyte-shaw-1994,piro-et-2006,bennett-et-2007,foster-giles-2008}\footnote{One of the authors (NW) personally experienced this phenomenon while living in Cambodia. On the day after major riots or other large political events, streets of the capital, Phnom Penh, were empty, most people stayed home out of fear of reprisal, and there was a general sense of apprehension.}. Similar to aspects of the previous two cases, we find dramatic behavioral changes the day after a large event and the changes we find are decreases in both call and mobility frequency.

{\bf Protests \--- November 19, 2008} (Figure S12, SI Appendix). Our system identified 45 sites in and extending well beyond Kigali that had unusual low call volume and movement frequency. Four additional sites recorded unusually low call volume. The sites are grouped in a large spatial cluster which is indicative of a major common cause of the disturbances at all these sites. On the same day, reports indicate that tens of thousands of Rwandans participated in a series of protests over the arrest in Germany of Rose Kabuye, a prominent Rwandan military and political figure, on alleged involvement in the plane crash that led to the 1994 genocide. The distance between the reported location of the protests (latitude -1.96, longitude 30.04) and the centroid of the closest site with unusual call volume and movement frequency is 1.8 km. Again, we find decreased call and mobility frequency, suggesting disruption in daily routines. In this case however, unlike the previous two cases, the behavioral anomalies occur on the same day as the event.

{\bf Floods \--- September 19, 2007} (Figure S13, SI Appendix). Our system identified 53 sites that had unusually low call volume and movement frequency on September 19, 2007. During the previous week, starting on September 19, torrential rains in the northwest of the country led to severe floods, leaving 15 people dead, 7000 people homeless and displaced, and more than 1000 houses uninhabitable. Floods also contaminated clean water supplies and decimated field crops, leading to concerns about waterborne diseases and food insecurity in the area. On September 18, floods dramatically swept away 42 homes and forced families to evacuate in the middle of the night. The following day, September 19, is when we find behavioral disruptions of decreased calling and movement. Notably, the behavioral anomalies occur across the country, instead of concentrated in the area most affected by flooding. In this case, the date of the behavioral disruption suggests a good match with the flooding event, but the spatial range of behavioral reaction decreases our confidence that the dramatic floods created the dramatic behavioral anomalies. It is possible that other areas of the country were also affected by flooding, that roads were damaged or transportation infrastructure was disrupted, or that families were busy rebuilding homes and crops that were destroyed by the rains. All of these possibilities are plausible explanations for decreased mobility and calling. However, further qualitative and quantitative research on behavioral reactions to similar flood disasters will be necessary to understand if and exactly how people change their communication and movement in response to natural disasters. The contribution of this case study is an indication that reactions to flood disasters might be much more complicated than we currently understand. 

{\bf Christmas Eve \--- December 24, 2007 and 2008} (Figures S14 and S15, SI Appendix). We identified 26 sites with unusually high call and movement frequency on December 24, 2007 and 59 such sites on December 24, 2008. Still more sites recorded only higher than usual call volume (21 in 2007 and 17 in 2008), or only higher than usual movement frequency (1 in 2007 and 2 in 2008). Given that about 90\% of Rwandans identify as Christians, it is not surprising that we find behavioral anomalies on Christmas Eve in 2007 and 2008. We expect that people called and visited their families to celebrate the holiday, resulting in the increases we find in both behaviors. The particular features of the behavioral anomaly we find on these two days (large spatial extent, higher than usual calls and mobility) match well the characteristics of this major planned religious event.

{\bf New Year's Eve and New Year's Day \--- January 1 and December 31, 2008, and January 1, 2009} (Figures S16, S17 and S18, SI Appendix). Our system identified more than 20 sites spread throughout Rwanda with unusually high call and movement frequency on each of January 1, 2008, December 31, 2008, and January 1, 2009. Given that New Year's is a national holiday that affects all people in Rwanda (regardless of religion) and given the wide spread of the behavioral anomalies we find, we believe that these anomalies are due to this holiday. Just as with Christmas, it is likely that Rwandans call and visit family and friends more often on New Year's Eve and Day.

{\bf International treaty \--- November 9, 2007} (Figure S19, SI Appendix). Behavioral anomalies were identified over a large area of Rwanda on November 9, 2007: 52 sites recorded unusually high call volume and movement frequency, three additional sites recorded unusually high call volume and one other site recorded unusually high movement frequency. One political event might explain this anomalous behavior: on that day, the governments of the Republic of Rwanda and of the Democratic Republic of Congo (DRC) signed the ``Nairobi Communiqu\'{e}" which defined a joint approach to end the threat to peace and stability in both countries and in the Great Lakes region posed by the Rwandan armed groups on Congolese territory. It is plausible that people made more calls to spread information and discuss this major treaty, but it is unclear why such as event would cause increased mobility. We do not find any other event that could plausibly have caused a nationwide response such as this.

{\bf Major unknown event \--- April 24 and 25, 2008} (Figures S20 and S21, SI Appendix). Our system identified unusually low call volume and movement frequency in 61 sites on April 24, 2008 and in 53 sites on the next day. On both days additional sites recorded unusually low call or movement frequency. We have been unable to find an event on or just before these days that could explain anomalous human behavior that lasted at least two consecutive days, affected almost the entire country and led to a significant decrease in the routine behaviors in Rwanda.

{\bf Commemoration of the genocide against the Tutsi \--- April 7 and 8, 2007, and April 7 and 8, 2008} (Figures S22, S23, S24 and S25, SI Appendix). Our system identified 26 sites with unusually low call volume and movement frequency on April 7, 2007 and 24 such sites on April 7, 2008. Our system also found a smaller number of sites with unusually low call volume and movement frequency on April 8, 2007 and 2008. April 7 is an official annual Rwandan holiday which marks the start date of the 1994 genocide. It is a planned event which affects most Rwandans. The behavioral anomalies spread across the country on these days for two years in a row suggest that the remembrance day could be the cause of decreased call volume and mobility frequency.

\section{Discussion}

In this paper, we contribute to the process of creating a system of detecting emergency events using mobile phone data. An effective event detection system could make significant contributions to humanitarian response and reducing the toll of disasters on human well-being. Towards this end, we develop a method for using mobile phone data to identify days with anomalous calling and mobility behavior, including days with high call volume and/or mobility, and low call volume and/or mobility. Our method also identifies the location of these anomalies and the geographical spread of the disturbances. We compare the days we identify with anomalous behaviors to a database of emergency and non-emergency events. Some days and places with behavioral anomalies match well with events and others do not. We learn from both cases.

Our analysis makes clear that detecting dramatic behavioral anomalies is only part of the work required to create an effective system of emergency event detection. The remaining work that is necessary is serious social-behavioral analysis of the exact types of behaviors that can be expected after different kinds of events and the exact time scales on which they occur. This will require intensive qualitative as well as quantitative analysis. It is only through a thorough understanding of these underlying differential behavioral patterns that an effective detection system can be developed.

This study reveals several dimensions of emergency events that must be considered for future work. We find that there are more days with anomalous decreases in calling and mobility than days with increases in these behaviors. Further, days with anomalous decreases in behavior match better with emergency events (including violence against civilians, protests, and a major flood), while days with increases in mobility and calling match better with joyous events, such as the Christmas and New Year's holidays. We find one irregularity in this pattern: the Lake Kivu earthquakes were followed by increased calling and mobility. Although our general finding of decreased behaviors after some threatening events contrasts common assumptions that people will be more likely to call and move about after emergencies, there are theoretical reasons to believe people will undertake these behaviors less often when busy responding to emergencies. It is also logically consistent that people will call and visit family and friends more during holidays. Consequently, examining decreases, as well as increases, in any behavior will likely yield key insights towards event detection.

We also find in this study different patterns of response to events for different behaviors. Here we examine call and mobility frequency. In some cases, both behaviors increase or decrease. In other cases, we find extreme increases in one behavior and extreme decreases in the other behavior at the same time and place. Other behaviors could also prove important in identifying events. Indeed, key insights will likely result from studying the particular combinations of increases and decreases of different behaviors, or the unique behavioral signatures of different events with various characteristics, dynamics, actors and causes. 

 Temporal patterns of behavior is another dimension that could be important in developing a better understanding of behavioral response to emergency events. The current version of our system is designed to detect anomalies on a daily basis. We were able to detect a wide range of events, from official holidays and the signing of international treaties to emergency events such as floods, violence against civilians or riots. But it is also possible that some responses occur within hours of an event. For example, people might call more often in the hour immediately following an event, then call less often for the rest of the day while they are busy responding to the event. As such, we would find different patterns if we examine calling behavior on an hourly versus a daily basis. 

Finally, examination of spatial patterns of response is also important. For some events, we find anomalies in responsive behaviors across large spaces, and for others we find that the area around a small number of cellular towers was affected. The spatial range of behavioral response is a key component of the unique behavioral signature of particular emergency and non-emergency events, and must be included in future research towards developing event detection systems. 

In summary, an effective system of emergency event detection, whether it uses CDRs, Twitter, or any other crowd sourced data, will be a result of close attention to detecting the exact signatures of human behaviors after different kinds of events. Currently, we know little about these exact signatures. Our analysis in this article suggests that these signatures are multi-dimensional and complex. In this situation, future progress on emergency event detection will require social scientific attention (quantitative and qualitative, theoretical and empirical) to human behavioral responses to emergency events. Our anomalous behavior detection system takes a step towards improving understanding of human responses to events, but this research is only the beginning. The only way this important, but difficult, task can be properly understood is through close multidisciplinary collaborations which involve social-behavioral scientists, statisticians, physicists, geographers and computer scientists.

\section*{Acknowledgments}

The authors thank Timothy Thomas and Matthew Dunbar for many useful discussions and for their help in processing GIS data. The authors are also grateful to Athena Pantazis and Joshua Rodd for recommending the Armed Conflict Location and Event Data Project. This work was partially supported by National Science Foundation Grant DMS-1120255 (to AD). NW's contribution has benefited from generous support of a NIH Pathways to Independence grant (R00HD067587). The funders had no role in study design, data collection and analysis, decision to publish, or preparation of the manuscript.

\appendix

\setcounter{figure}{0}
\renewcommand{\figurename}{Figure }
\renewcommand{\thefigure}{S\arabic{figure}}

\renewcommand{\tablename}{Table }
\renewcommand{\thetable}{S\arabic{table}}

\section*{Supplementary Information (SI) Appendix}

\section*{SI1: The road network and the grid cell system}

\begin{figure}[htbp!]
\begin{center}
 \includegraphics[height=5.5in,angle=0]{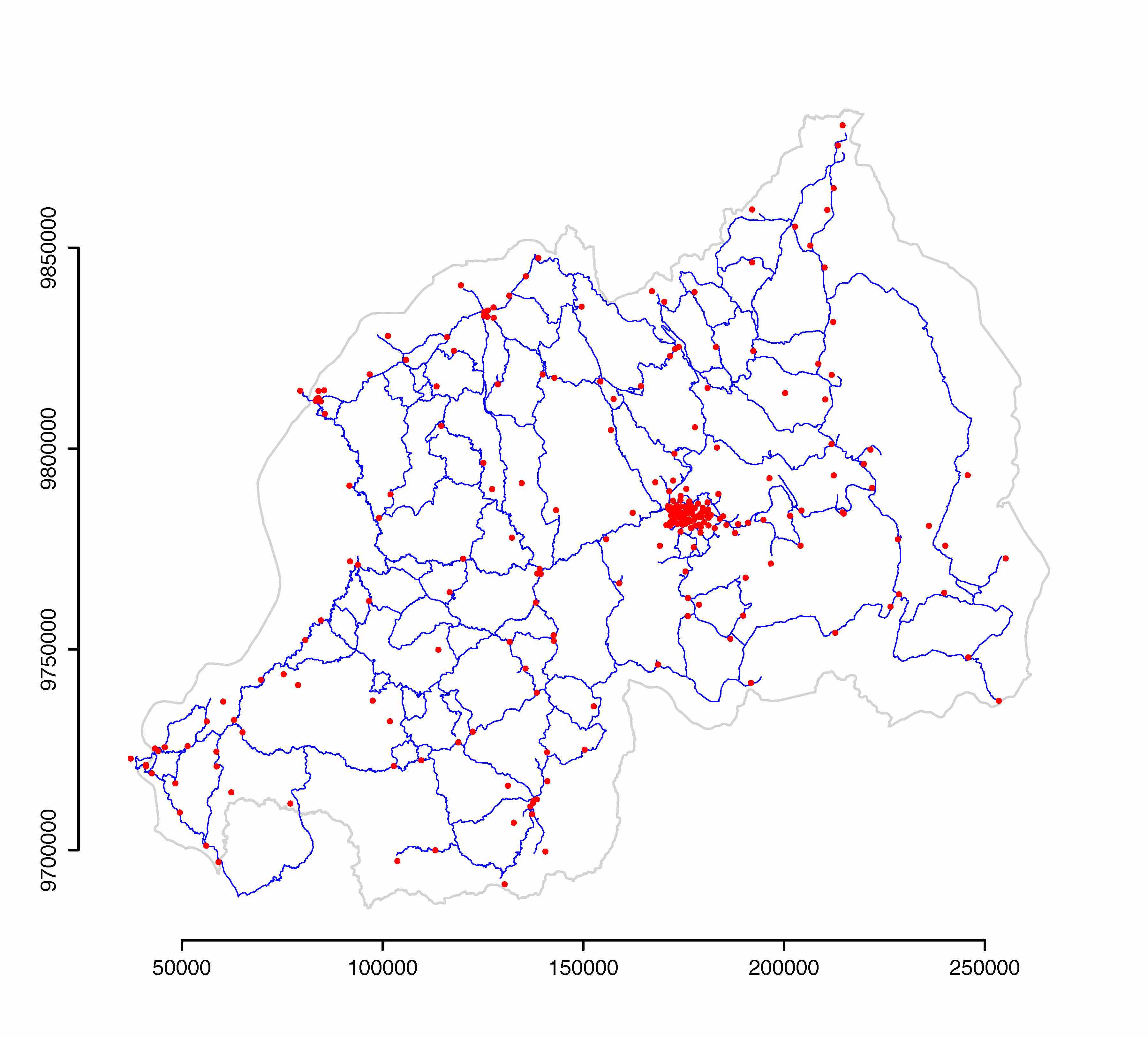}
 \caption{\label{fig:towersRoads}{\bf Rwandan road network system}. Map of Rwanda showing the position of the 269 cellular towers (red) and the structure of the network of roads that are also segments in quickest routes (blue).}
\end{center}
\end{figure}

\begin{figure}[htbp!]
\begin{center}
    \includegraphics[height=4.5in,angle=0]{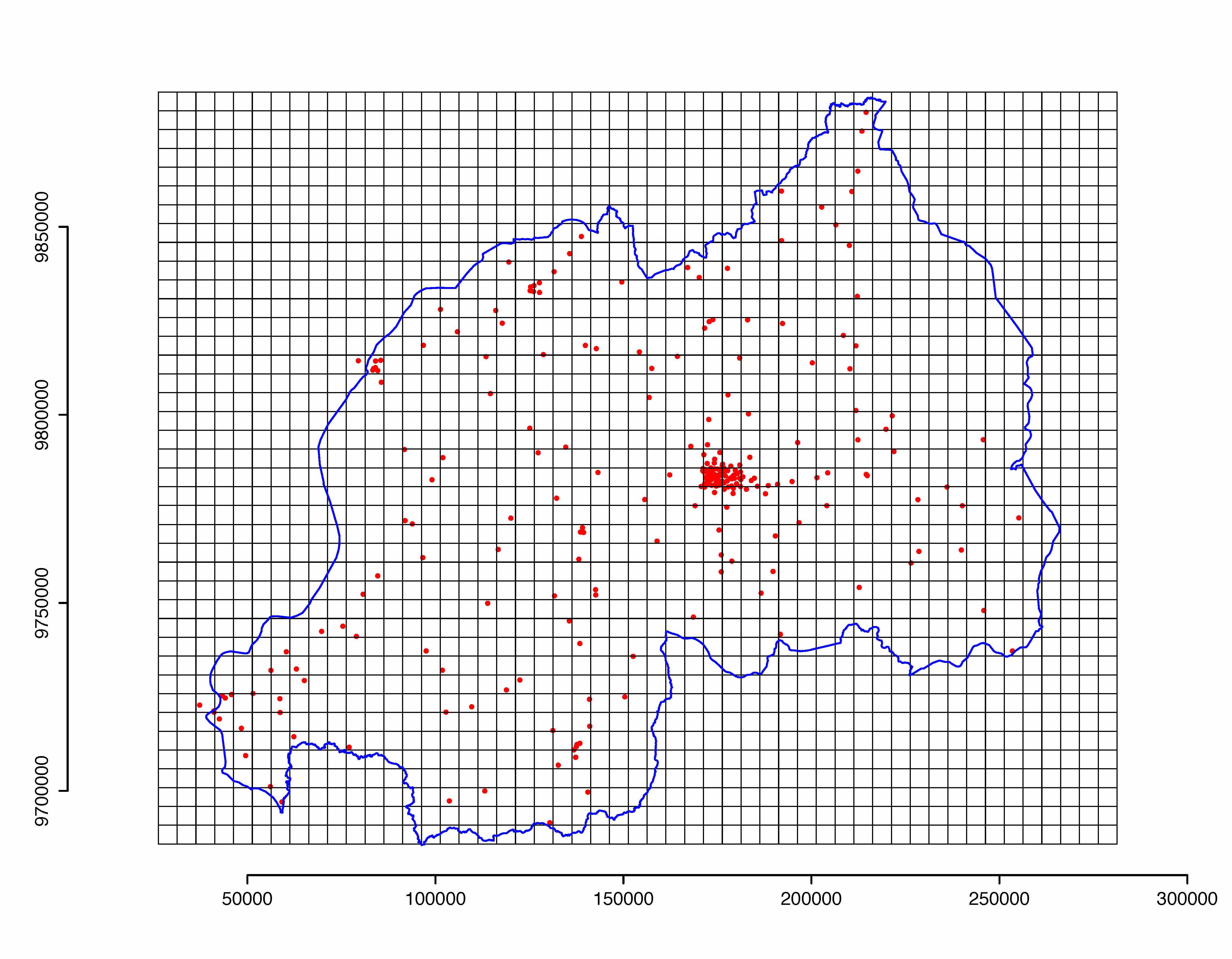}
 \caption{\label{fig:towersGridcells}{\bf Rwandan grid cell system}. Map of Rwanda showing the position of the cellular towers (red) with respect to the 2040 5 km x 5 km grid cells. Rwanda's boundary is shown in blue.}
\end{center}
\end{figure}

The methodology presented in this paper is based on two Geographic Information Systems (GIS) components: a road network system for Rwanda and a grid cell system which divides a spatial bounding box for Rwanda's boundary into 2040 5 km x 5 km cells. Figures \ref{fig:towersRoads} and \ref{fig:towersGridcells} display the locations of the $269$ cellular towers that appear in the Rwandan CDR data with respect to the road network and the grid cell system, respectively. The grid cells that contain at least one tower are called sites. Only 155 out of the 2040 grid cells are sites. Four sites in the Kigali area contain the largest number of cellular towers: 41, 22, 6 and 5, respectively. Seven sites contain four towers, four sites contain three towers, 14 sites contain two towers and the other sites contain only one tower. These counts represent the towers that belong to a site between June 1, 2005 and January 1, 2009. In any period of time between these dates, all, some or none of the towers that belong to a site are actually active (i.e. handle cellular communications). As such, the number of sites (i.e., grid cells that contain at least one active tower) at any time might be smaller than 155.

\subsection*{SI1.1: The Road Network System} 

We use road network data from the crowd sourced OpenStreetMap\footnote{\url{http://www.openstreetmap.org/}}. Roads are categorized with respect to their quality in the following hierarchy: trunk roads, primary roads, secondary roads and tertiary roads. We estimate that the average speeds of travel for these four types of roads are 120 km/h, 60 km/h, 45 km/h and 30 km/h, respectively. Based on this determination, we employed ESRI's ArcGIS\footnote{\url{http://www.esri.com/software/arcgis}} to determine approximate travel distances and travel times between the centroids of pairs of sites. We used the function ``Closest Facility''  of ArcGIS Network Analyst\footnote{\url{http://www.esri.com/software/arcgis/extensions/networkanalyst}} to identify the quickest road paths between the centroids of any pair of sites and stored these ${155 \choose 2 } = 23870$ routes together with their corresponding travel distances and travel times. We also identify the sites on the quickest route between the centroids of each pair of sites.

\subsection*{SI1.2: The Grid Cell System}

We overlay a customized rectangular grid with square cells of equal size on the map of Rwanda, and replace cellular tower locations with the centroid of the sites they belong to. Instead of measuring straight line distances from tower to tower, we measure distances between the centroids of the sites via the quickest road route which connects these centroids. The raw road network data downloaded from OSM was such that 11 sites were not intersected by the Rwandan road network. To connect these sites to the road network, we moved the location of their centroids to adjacent grid cell centroids. 

Choosing the size of the grid cells is an important decision. Based on geographical and technological considerations, we estimated catchment areas in which a user of a cellular tower is likely to be located. We estimate that the maximum signal distance for the type of towers in Rwanda is roughly 10 km. Several factors further reduce this maximum signal distance, including relative location of a user with respect to a tower, topography of the areas surrounding towers, and the decay in signal strength with increasing distances from towers. As such, we  reduce the maximum user-to-tower distance to 5 km. The resulting 5 km x 5 km grid cell system is a 51 x 40 matrix (2040 grid cells) that covers 51,000 km$^2$ extending just outside of the border of Rwanda \--- see Figure \ref{fig:towersGridcells}. Each grid cell is indexed by a number from 1 to 2040: grid cell 1 is located in the lower left corner and grid cell 2040 is located in the upper right corner. The indices increase first by row, then by column. Each of the 269 cellular towers is subsequently mapped to its corresponding grid cell (site).

\section*{SI2: Measures of human behavior}

Consider the sequence of CDRs associated with a mobile phone in a reference period of time $\mathcal{T}$ (e.g., a day, a week, a month or a year):
\begin{eqnarray} 
 M = \{ m_1,m_2,\ldots,m_n\}.\label{eq:cdrSeq}
\end{eqnarray}
We assume that the wireless-service provider that generated these CDRs has $K$ active towers in the reference time period $\mathcal{T}$, and that the spatial locations $l^{CT}_i$, $i \in \mathcal{K}=\{1,2,\ldots,K\}$ of these active towers are known. In (\ref{eq:cdrSeq}), $m_i\in \mathcal{K}$, $1\le i\le n$, is the identifier of the cellular tower that handled the communication represented by the $i$-th CDR in the sequence. If $i<j$ the communication represented by $m_i$ was recorded before the communication represented by $m_j$.  We refer to $M$ as the spatiotemporal trajectory of the cellular phone that generated the sequence of CDRs. We assume that the region of interest was divided into non-overlapping grid cells identified by indices in $\mathcal{Q}\in \{1,2,\ldots,Q\}$. We denote by $l^{GC}_j$ the location of the centroid of the grid cell $j\in \mathcal{Q}$. We introduce a mapping function $q^{GC}(\cdot)$ which gives, for each cellular tower $i\in \mathcal{K}$, the grid cell $q^{GC}(i)\in \mathcal{Q}$ the tower belongs to. The sites are those grid cells that contain at least one tower:
$$
 \mathcal{S} = \left\{j: j\in \mathcal{Q} \mbox{ such that there exists }  i\in \mathcal{K} \mbox{ with } q^{GC}(i)=j\right\}.
$$
Since we assume that all the towers indexed by $\mathcal{K}$ are active in the reference time period $\mathcal{T}$, $\mathcal{S}$ represents the set of sites in $\mathcal{T}$.

We transform the spatiotemporal trajectory $M$ from (\ref{eq:cdrSeq}) into the corresponding time ordered sequence of sites to which the active towers that appear in $M$ belong to:
\begin{eqnarray}
 M^{GC} = \{ g_1,g_2,\ldots,g_n\}, \label{eq:gridSeq}
\end{eqnarray}
\noindent where $g_i = q^{GC}(m_i)\in \mathcal{S}$.

The measure of behavior called ``call volume" is the number of times a person communicates in the reference time period $\mathcal{T}$. For the spatiotemporal trajectory $M^{GC}$, this measure is equal with the length of the sequence $n$.

The measure of behavior called ``movement frequency" (also referred to as ``number of trips" in \cite{williams-et-2014}) is a count of the number of times a person communicates from a different grid cell than their previous communication:
$$
 \# \left\{ i: i\in \{1,2,\ldots,n-1\} \mbox{ such that } g_i\ne g_{i+1}\right\}.
$$

To see why the movement frequency captures an aspect of human behavior complementary to the call volume, consider an example person who makes 10 calls from one site and another example person that calls once from 10 different sites. The call volumes of the two persons are equal. But the movement frequency of the first person is 0, while the movement frequency of the second person is 9. The behaviors of these two persons are dissimilar: the first one makes multiple calls from one site and is not mobile, while the second person moves significantly more. In this paper we employ only one measure of mobility, but combinations of several measures of mobility can also be explored. See \cite{williams-et-2014} for an in-depth discussion of measures of mobility constructed from mobile phone records.

\section*{SI3: Identifying days with anomalous human behavior at one site}

We consider a reference time period P of T consecutive days.  We denote by $Y_{i,t}$ the behavioral measurement associated with the $i$-th person that made at least one call from a site S during day $t$ with $i\in \{1,2,\ldots,n_t\}$. Since $Y_{i,t}$ represent either counts of the number of calls or counts of the number of trips, we assume Poisson sampling models for measurements within each day:
$$
 Y_{1,t},Y_{2,t},\ldots,Y_{n_t,t}\mid \theta_t \sim \mbox{ i.i.d. } \Poi(\theta_t).
$$
The Poisson means $\{\theta_t: t\in \mbox{P}\}$ have independent Gamma prior distributions $\Gam(a,b)$. The shape parameter $a$ and the rate parameter $b$ are set to $1$, and yield proper priors with mode equal to $0$, and with mean and variance equal with $1$. Therefore, a priori, we assume that individuals make 1 call and make 1 trip in any given day. The posterior distribution of $\theta_t$ is also Gamma:
$$
 \theta_t\mid Y_{1,t},Y_{2,t},\ldots,Y_{n_t,t} \sim \Gam(a+\sum_{i=1}^{n_t}Y_{i,t},b+n_t).
$$
The rate parameter $b$ is interpreted as the number of prior observations. Given that every day hundreds or thousands of people make calls from each site, the behavioral measurements associated with each day and each site have a large weight in the posterior distributions of the Poisson means $\theta_t$ with respect to the Gamma priors. Predictions about the measurement $\tilde{Y}_t$ of a new person that calls from site S during day $t$ which account for uncertainty about the Poisson means are obtained based on the predictive distribution of $\tilde{Y}_t$ that is a negative binomial
\begin{eqnarray}
 \tilde{Y}_t \mid Y_{1,t},Y_{2,t},\ldots,Y_{n_t,t} \sim \negbin (a+\sum_{i=1}^{n_t}Y_{i,t},b+n_t).
 \label{eq:prednegbin}
\end{eqnarray}
A Monte Carlo estimate of the probability that a random caller from day $t_0$ in the time period P had a larger behavioral measurement (i.e., made more calls or moved more frequently) than a random caller from a random day in P other than $t_0$ is obtained by repeating the following steps $N=10000$ times. We work with a counter $l_{t_0}$ initialized at $0$.
\begin{enumerate}
 \item For each day $t$ in the reference time period P simulate $\tilde{Y}_t$ from the negative binomial predictive distribution (\ref{eq:prednegbin}).
 \item Sample a day $t_1$ in $\mbox{P}\setminus\{t_0\}$.
 \item If $\tilde{Y}_{t_0}>\tilde{Y}_{t_1}$ increment $l_{t_0}$ by 1.
\end{enumerate}
The Monte Carlo estimate is given by $l_{t_0}/N$.

\begin{figure}[htbp!]
\begin{center}
\includegraphics[width=4.5in,angle=0]{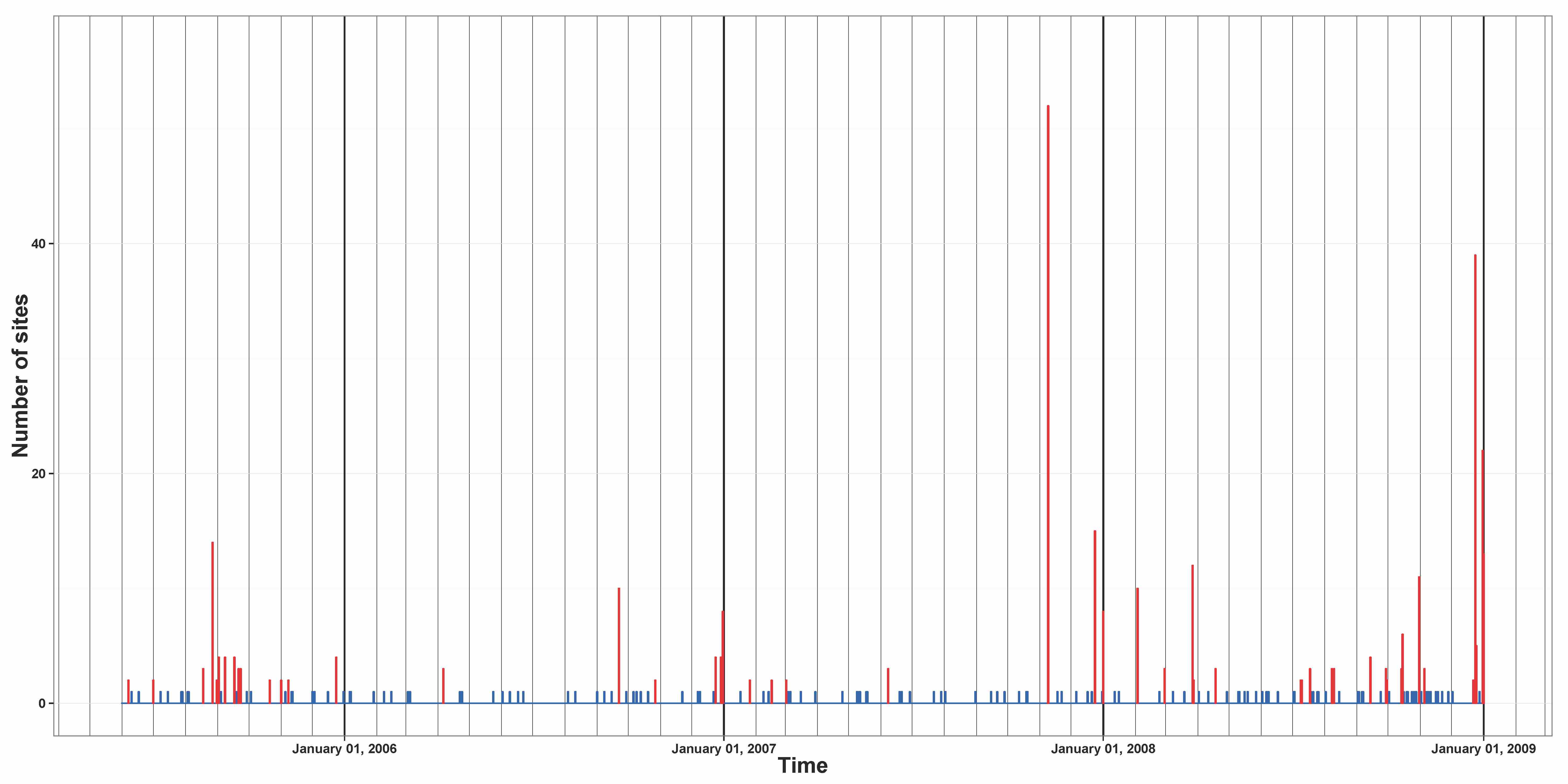}
\caption{{\bf Daily time series of the size of the largest spatial clusters of sites with high call volume and movement frequency.} For each day, we show the maximum number of sites that belong to the same spatial cluster and recorded higher than usual call volume and movement frequency. Red indicates days for which the largest spatial cluster has at least two sites, while blue indicates days for which the largest spatial cluster has one site.}
\label{fig:UPTimeSeriescallsmovement}
\end{center}
\end{figure}

\begin{figure}[htbp!]
\begin{center}
\includegraphics[width=4.5in,angle=0]{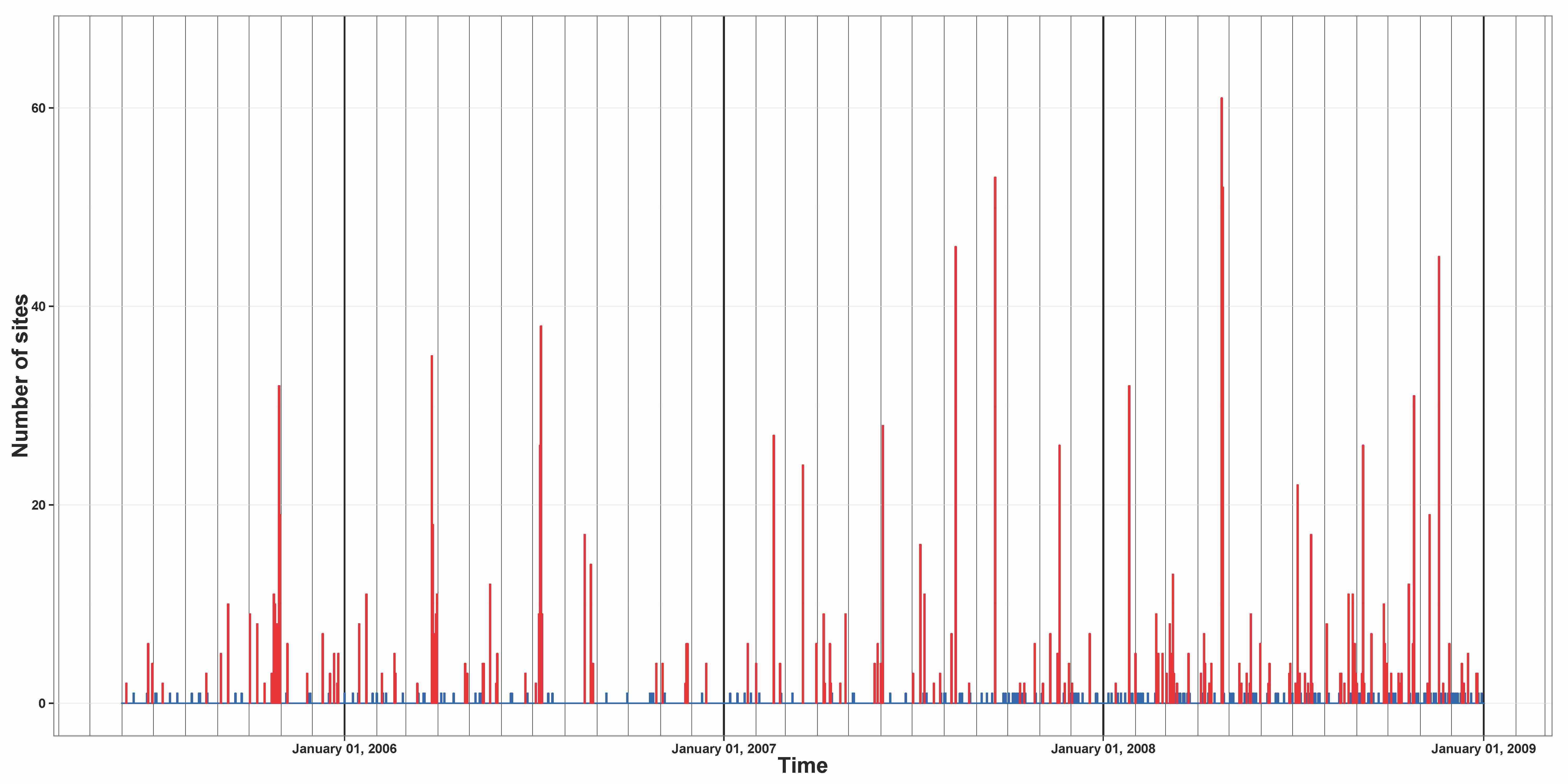}
\caption{{\bf Daily time series of the size of the largest spatial clusters of sites with low call volume and movement frequency.} For each day, we show the maximum number of sites that belong to the same spatial cluster and recorded lower than usual call volume and movement frequency. Red indicates days for which the largest spatial cluster has at least two sites, while blue indicates days for which the largest spatial cluster has one site.}
\label{fig:DOWNTimeSeriescallsmovement}
\end{center}
\end{figure}

\begin{figure}[htbp!]
\begin{center}
\includegraphics[width=4.5in,angle=0]{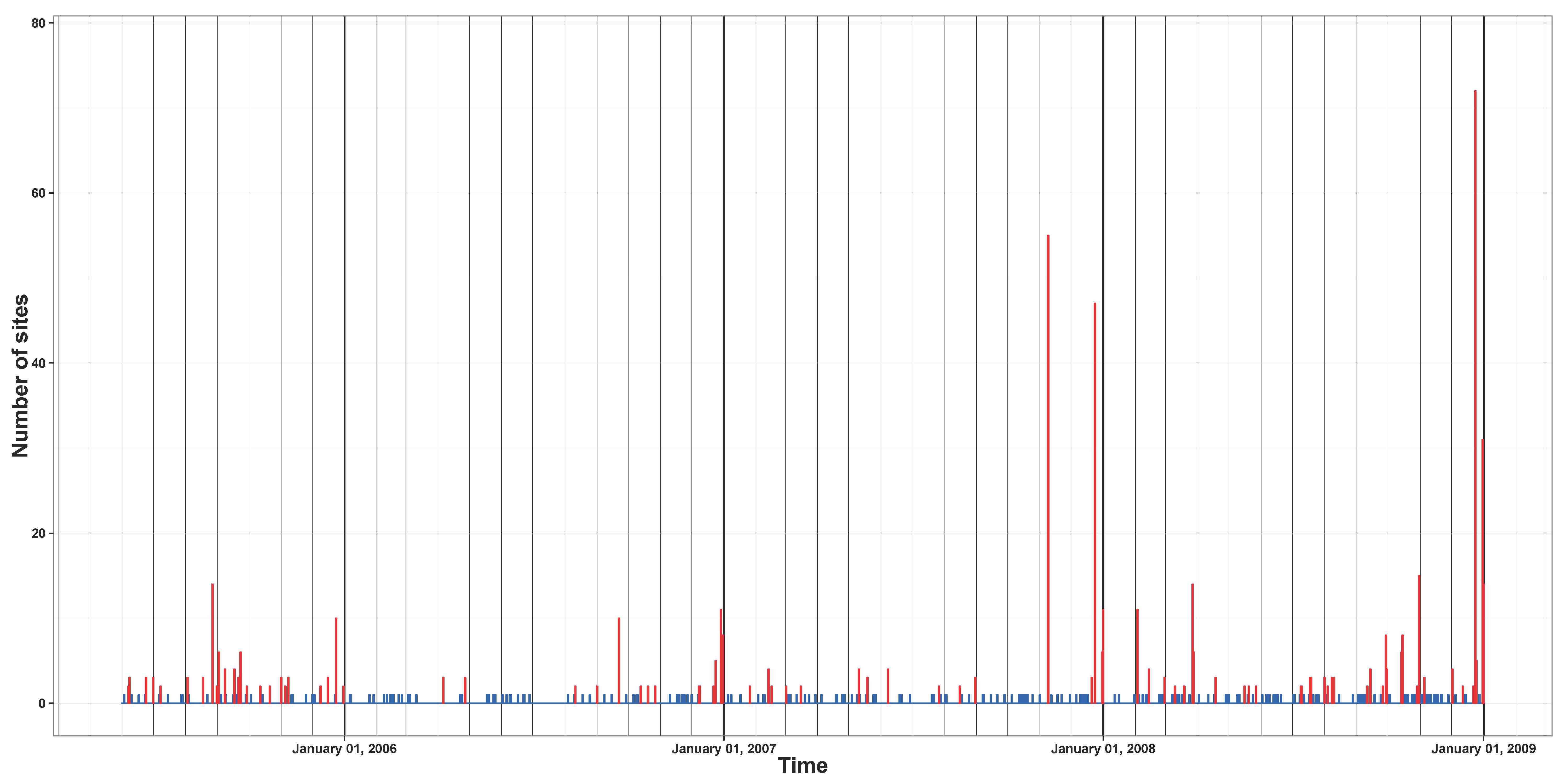}
\caption{{\bf Daily time series of the size of the largest spatial clusters of sites with high call volumes.} For each day, we show the maximum number of sites that belong to the same spatial cluster and recorded higher than usual call volume. Red indicates days for which the largest spatial cluster has at least two sites, while blue indicates days for which the largest spatial cluster has one site.}
\label{fig:UPTimeSeriescalls}
\end{center}
\end{figure}

\begin{figure}[htbp!]
\begin{center}
\includegraphics[width=4.5in,angle=0]{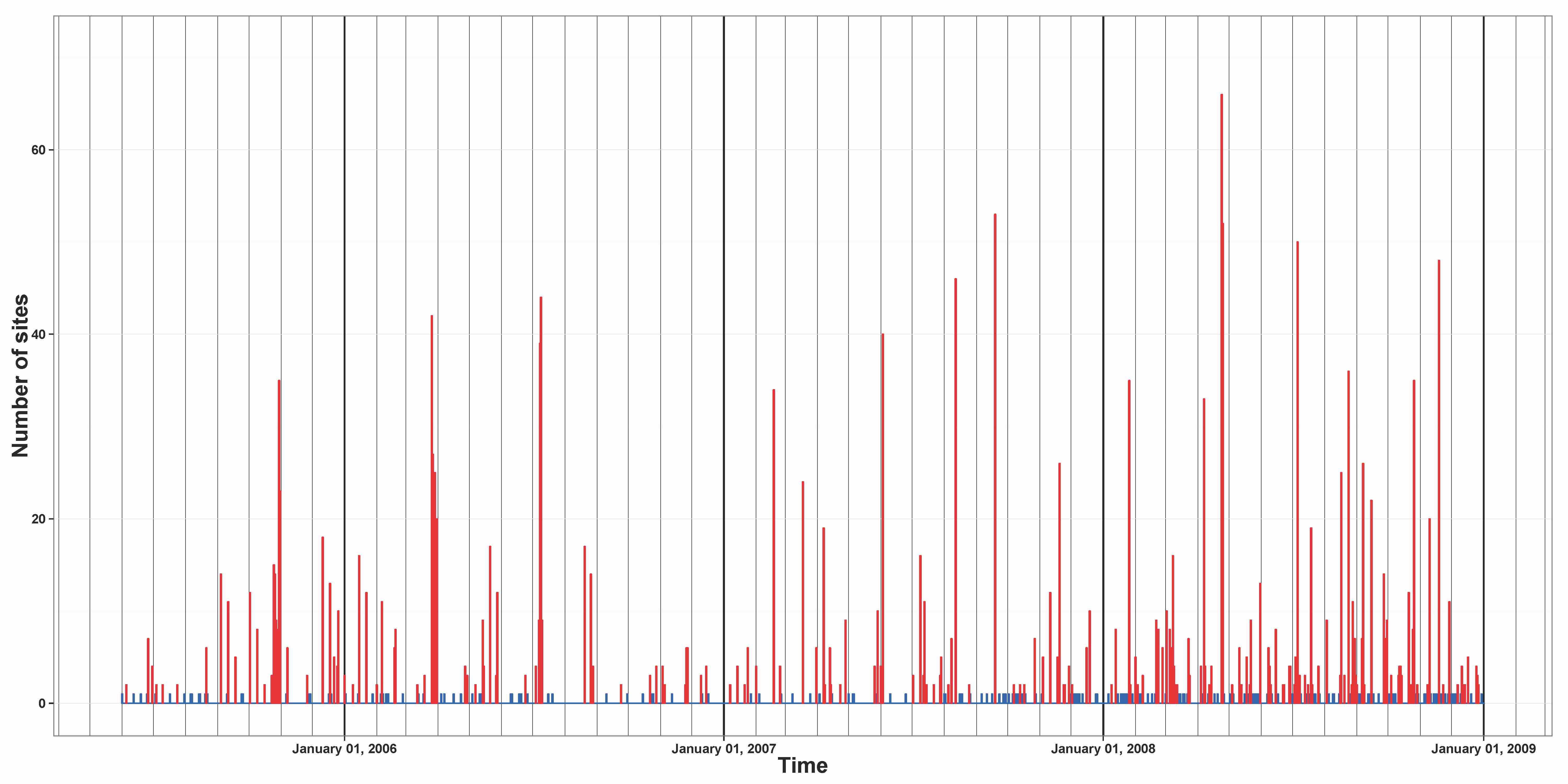}
\caption{{\bf Daily time series of the size of the largest spatial clusters of sites with low call volume.} For each day, we show the maximum number of sites that belong to the same spatial cluster and recorded lower than usual call volume. Red indicates days for which the largest spatial cluster has at least two sites, while blue indicates days for which the largest spatial cluster has one site.}
\label{fig:DOWNTimeSeriescalls}
\end{center}
\end{figure}

\begin{figure}[htbp!]
\begin{center}
\includegraphics[width=4.5in,angle=0]{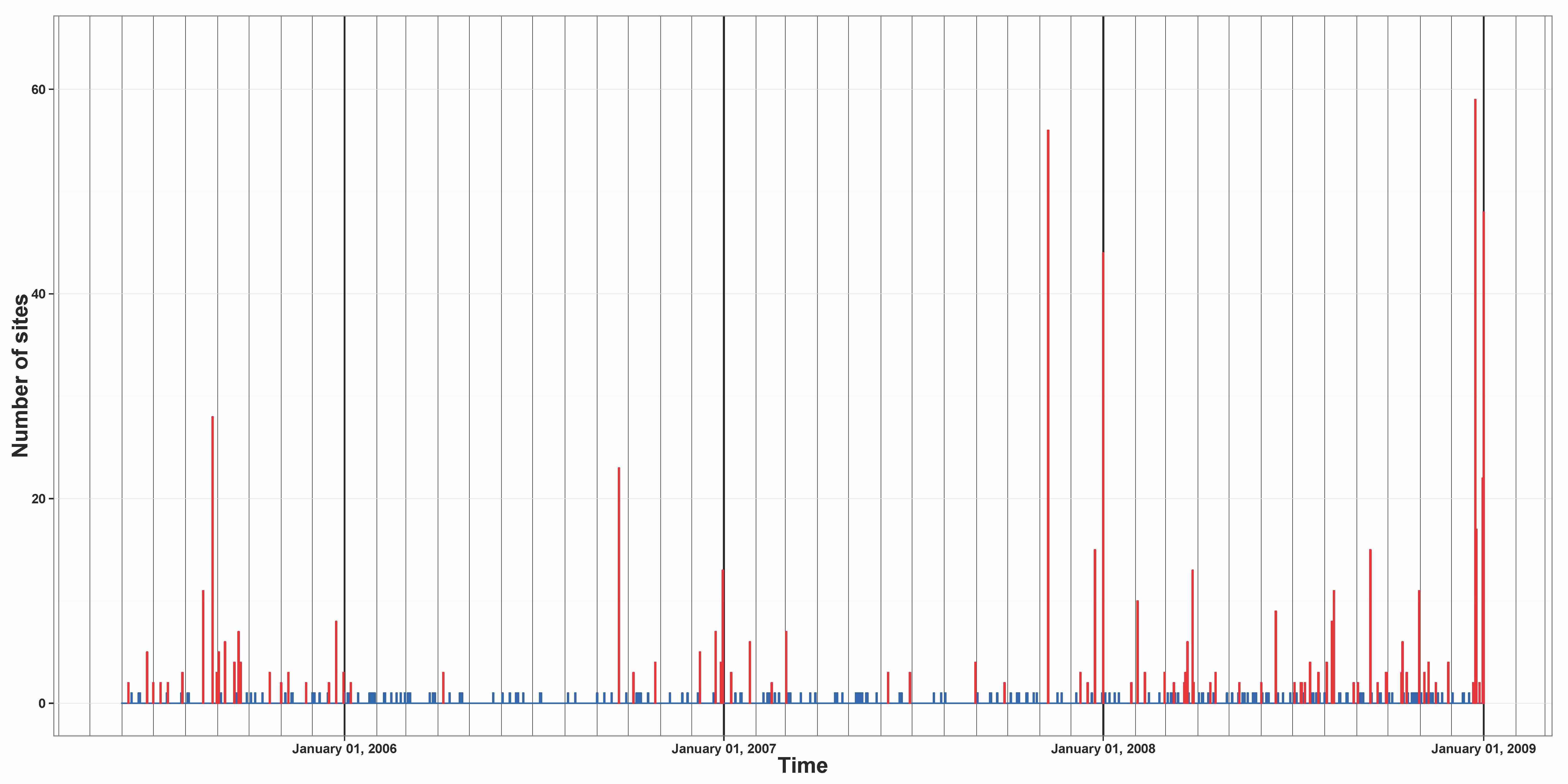}
\caption{{\bf Daily time series of the size of the largest spatial clusters of sites with high movement frequencies.} For each day, we show the maximum number of sites that belong to the same spatial cluster and recorded higher than usual movement frequency. Red indicates days for which the largest spatial cluster has at least two sites, while blue indicates days for which the largest spatial cluster has one site.}
\label{fig:UPTimeSeriesmovement}
\end{center}
\end{figure}

\begin{figure}[htbp!]
\begin{center}
\includegraphics[width=4.5in,angle=0]{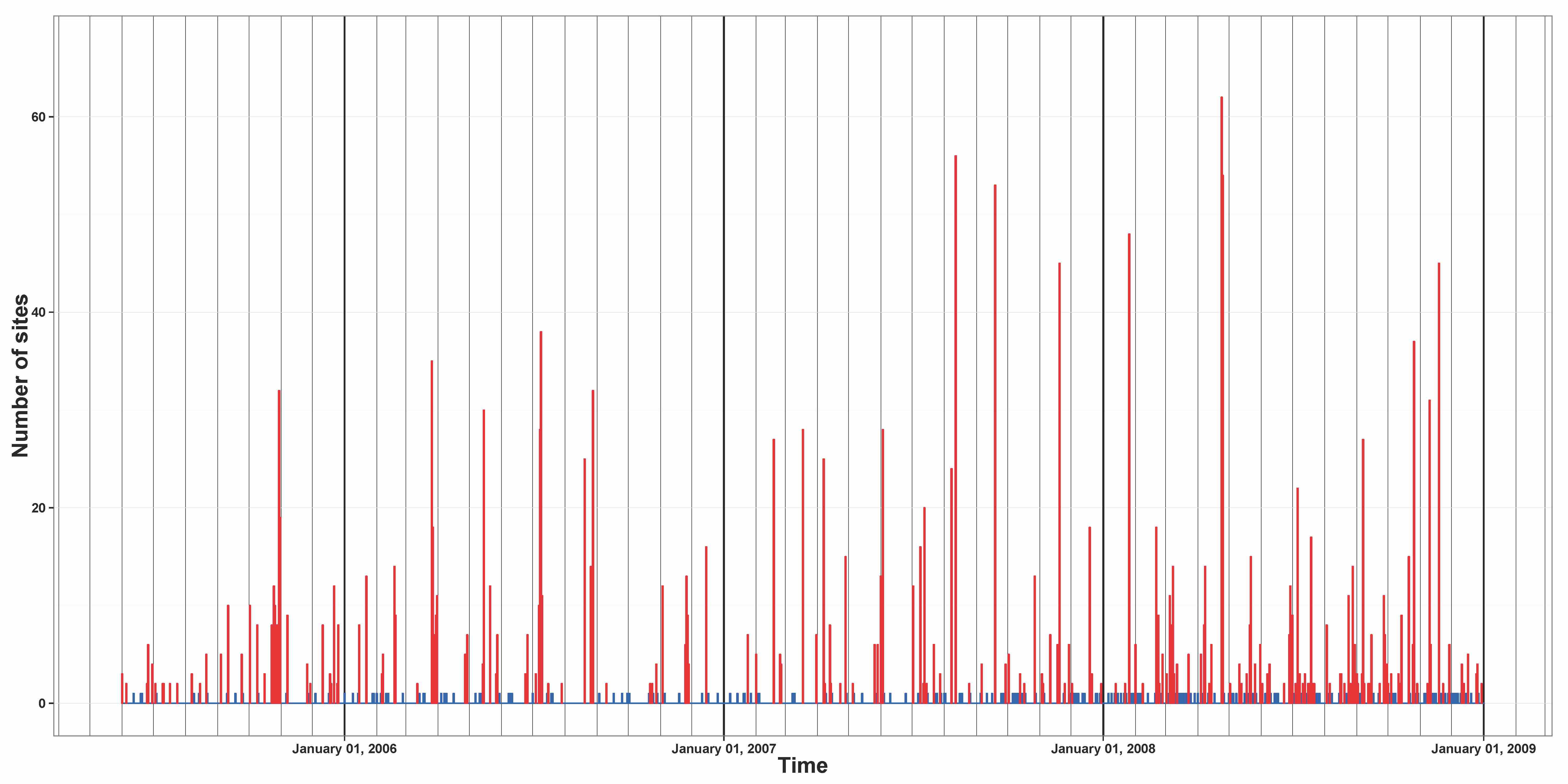}
\caption{{\bf Daily time series of the size of the largest spatial clusters of sites with low movement frequency.} For each day, we show the maximum number of sites that belong to the same spatial cluster and recorded lower than usual movement frequency. Red indicates days for which the largest spatial cluster has at least two sites, while blue indicates days for which the largest spatial cluster has one site.}
\label{fig:DOWNTimeSeriesmovement}
\end{center}
\end{figure}

\begin{figure}[htbp!]
\begin{center}
\includegraphics[width=5.5in,angle=0]{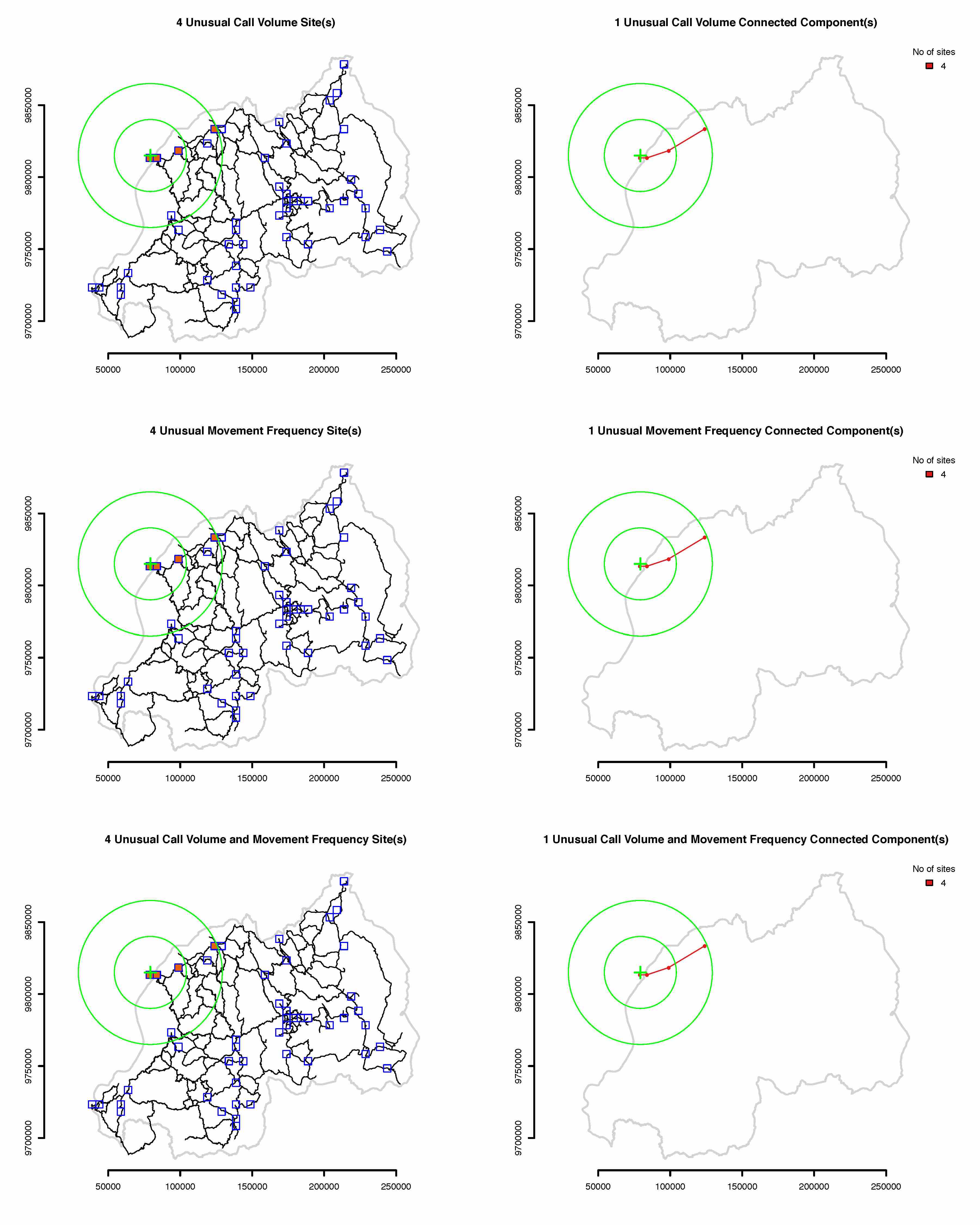}
\caption{{\bf Sites with unusually high behavior on September 17, 2005.}  Four sites recorded unusually high call volume and movement frequency, and belong to the same spatial cluster. This anomalous pattern of communication was potentially caused by an ACLED event recorded on September 16, 2005. The green cross marks the reported location of this event, while the two green circles mark the 25 and 50 km areas around this location. The distance between the event's location and the centroid of the closest site with unusual communication activity is 1.7 km.}
\label{fig:UPSeptember172005}
\end{center}
\end{figure}

\begin{figure}[htbp!]
\begin{center}
\includegraphics[width=5.5in,angle=0]{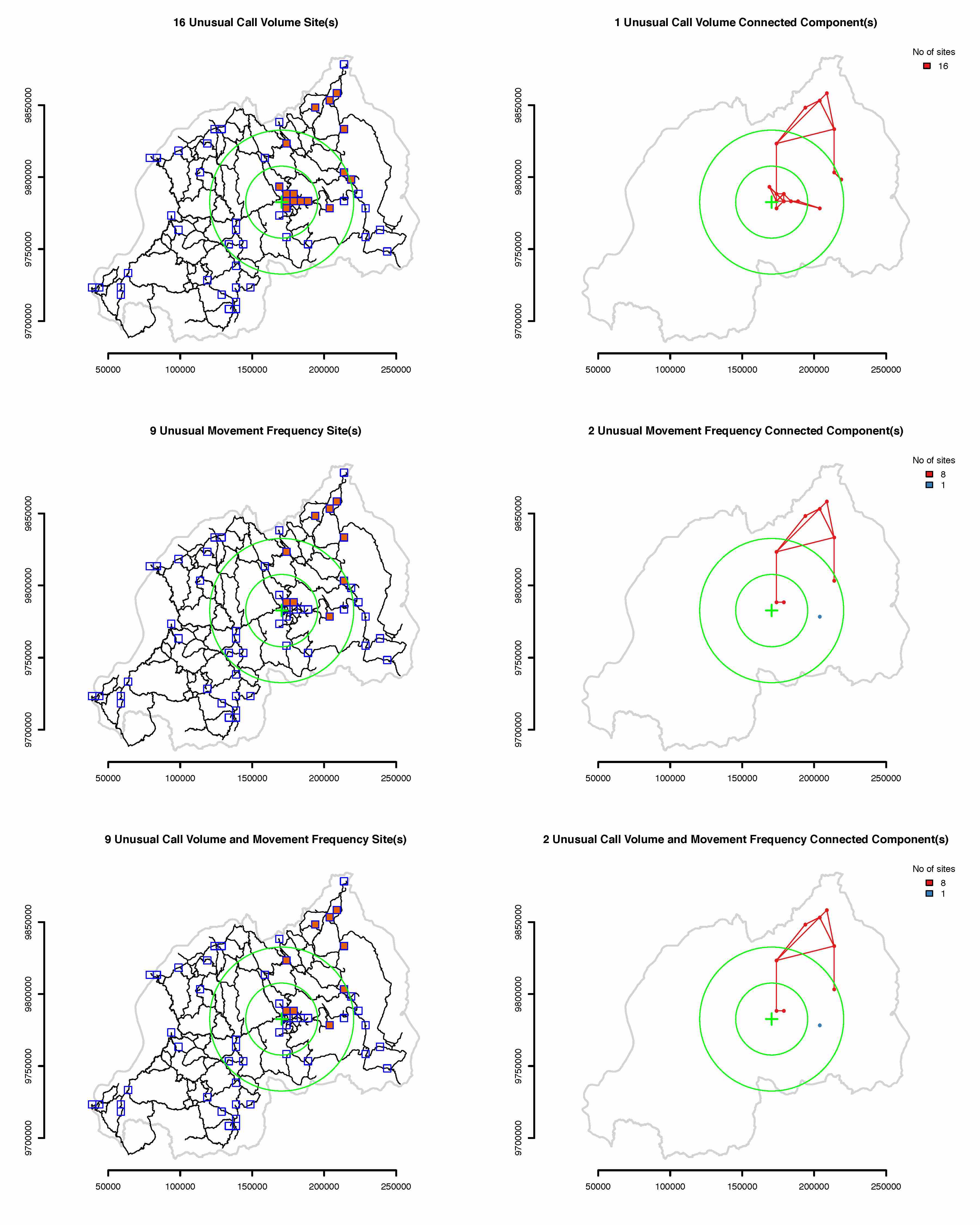}
\caption{{\bf Sites with unusually low behavior on January 15, 2006.} Nine sites recorded unusually low call volume and movement frequency. Seven additional sites recorded unusually low call volume. Most of the sites in these two groups belong to one spatial cluster. One spatial cluster with only one site is also present. This anomalous pattern of communication was potentially caused by an ACLED event recorded on the same day. The green cross marks the reported location of this event, while the two green circles mark the 25 and 50 km areas around this location. The distance between the event's location and the centroid of the closest site with unusual call volume (movement frequency) is 3.8 (5.4) km.}
\label{fig:DOWNJanuary152006}
\end{center}
\end{figure}

\begin{figure}[htbp!]
\begin{center}
\includegraphics[width=5.5in,angle=0]{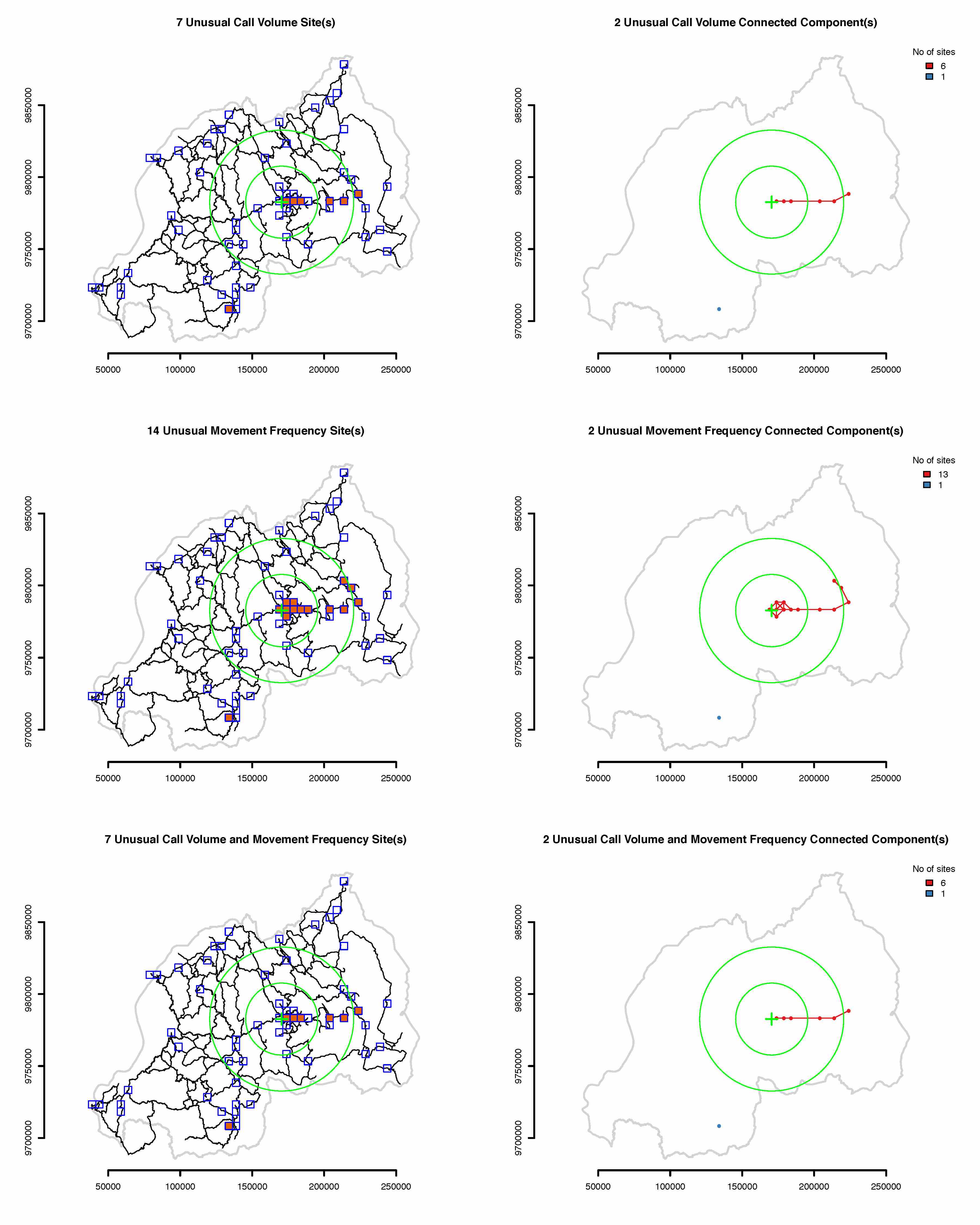}
\caption{{\bf Sites with unusually low behavior on November 26, 2006.} Seven sites recorded unusually low call volume and movement frequency. Seven additional sites recorded unusually low movement frequency. Most of the sites in these two groups belong to one spatial cluster. One spatial cluster with only one site is also present. This anomalous pattern of communication was potentially caused by an ACLED event recorded on November 25, 2006. The green cross marks the reported location of the event, while the two green circles mark the 25 and 50 km areas around this location. The distance between the event's location and the centroid of the closest site with unusual call volume (movement frequency) is 3.38 (1.83) km.}
\label{fig:DOWNNovember262006}
\end{center}
\end{figure}

\begin{figure}[htbp!]
\begin{center}
\includegraphics[width=5.5in,angle=0]{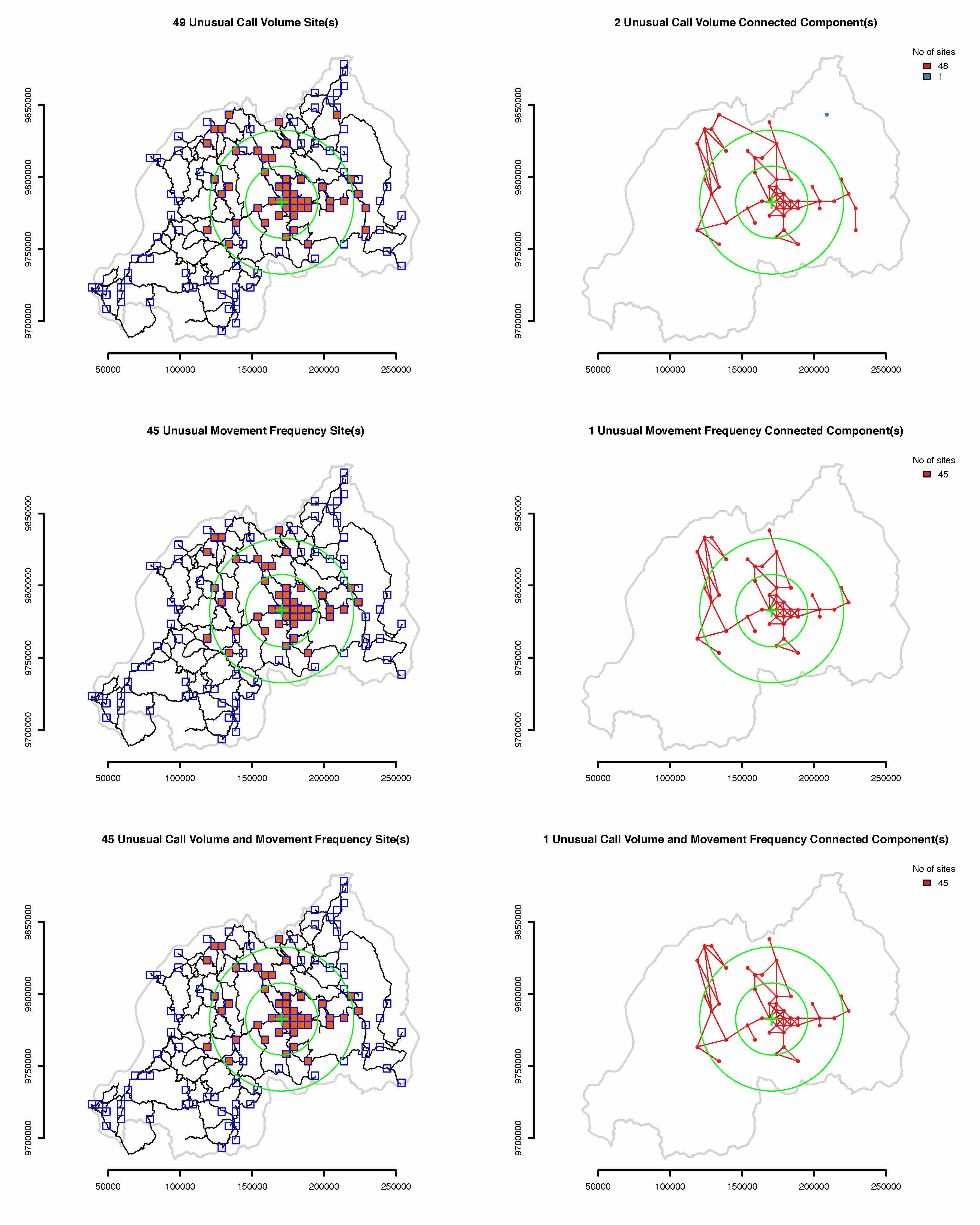}
\caption{{\bf Sites with unusually low behavior on November 19, 2008.} A number of 45 sites recorded unusually low call volume and movement frequency. Four additional sites recorded unusually low call volume. The sites in these two groups belong to one spatial cluster. This anomalous pattern of communication was potentially caused by an ACLED event recorded on the same day. The green cross marks the reported location of this event, while the two green circles mark the 25 and 50 km areas around this location. The distance between the event's location and the centroid of the closest site with unusual call volume and movement frequency is 1.8 km.}
\label{fig:DOWNNovember192008}
\end{center}
\end{figure}

\begin{figure}[htbp!]
\begin{center}
\includegraphics[width=5.5in,angle=0]{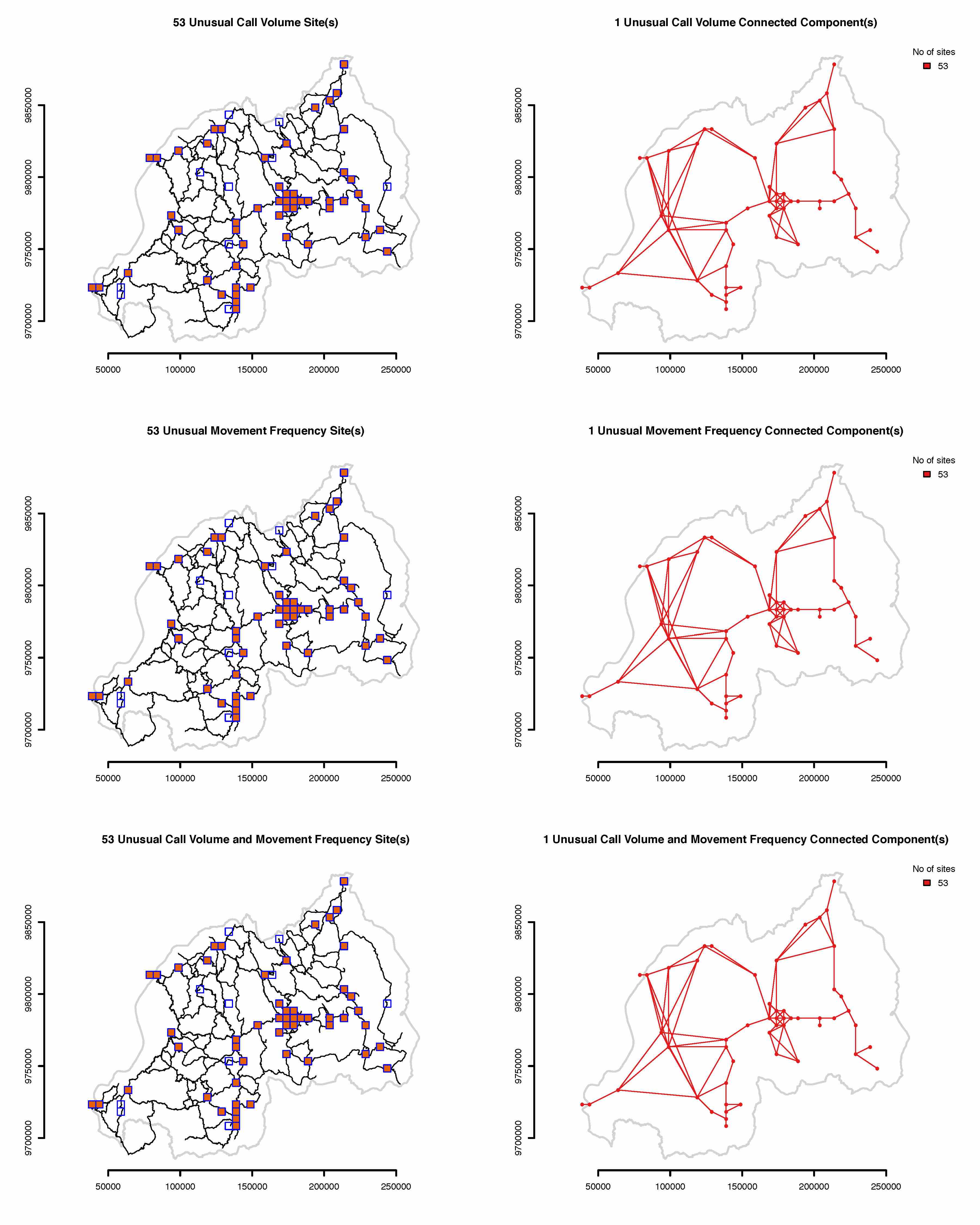}
\caption{{\bf Sites with unusually low behavior on September 19, 2007.}  A number of 53 sites recorded unusually low call volume and movement frequency. These sites belong to the same spatial cluster.}
\label{fig:DOWNSeptember192007}
\end{center}
\end{figure}

\begin{figure}[htbp!]
\begin{center}
\includegraphics[width=5.5in,angle=0]{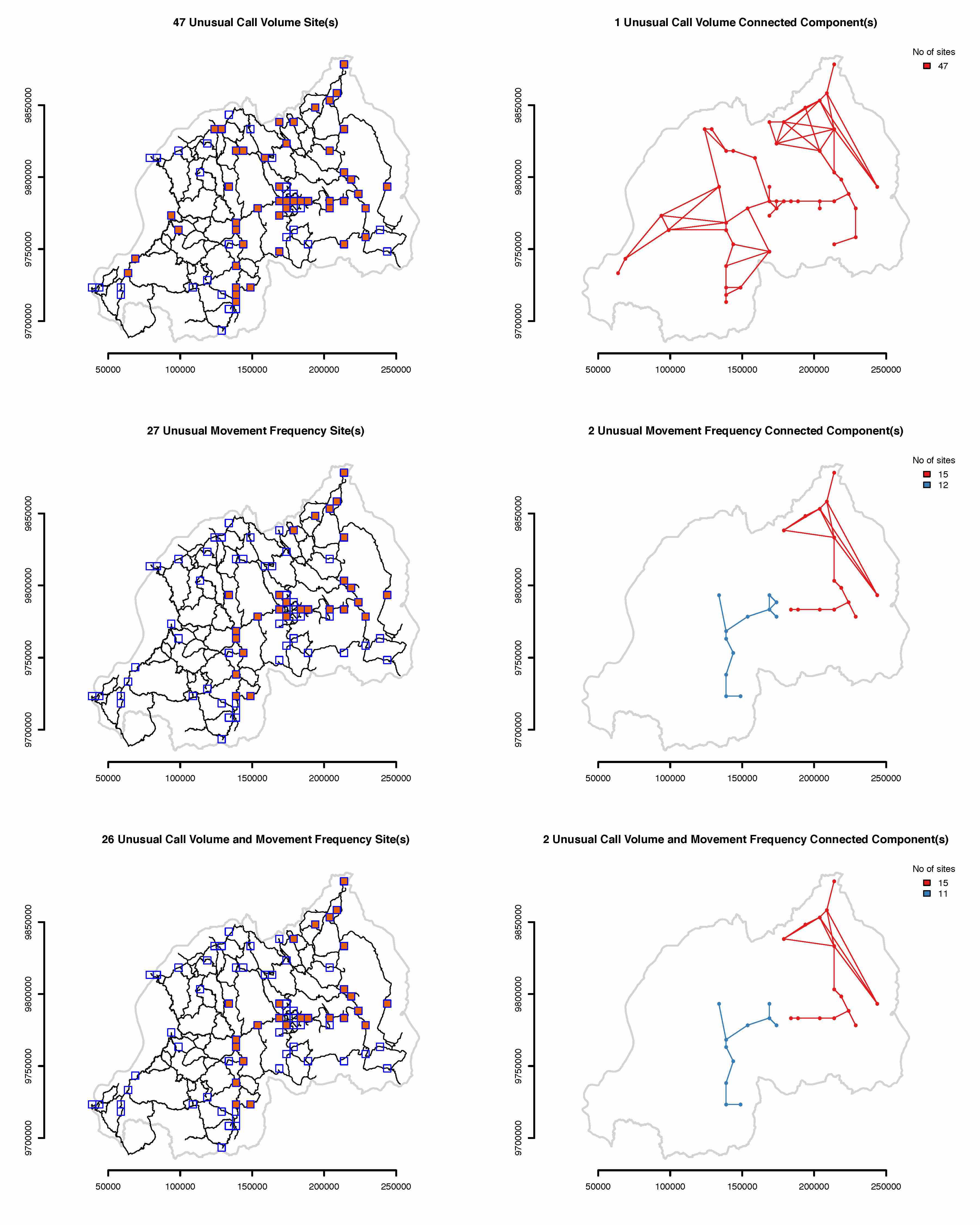}
\caption{{\bf Sites with unusually high behavior on December 24, 2007.}  A number of 26 sites recorded unusually high call volume and movement frequency. A number of 21 additional sites recorded unusually high call volume, while one other site recorded unusually high movement frequency. The sites in these three groups belong to one or two spatial clusters.}
\label{fig:UPDecember242007}
\end{center}
\end{figure}

\begin{figure}[htbp!]
\begin{center}
\includegraphics[width=5.5in,angle=0]{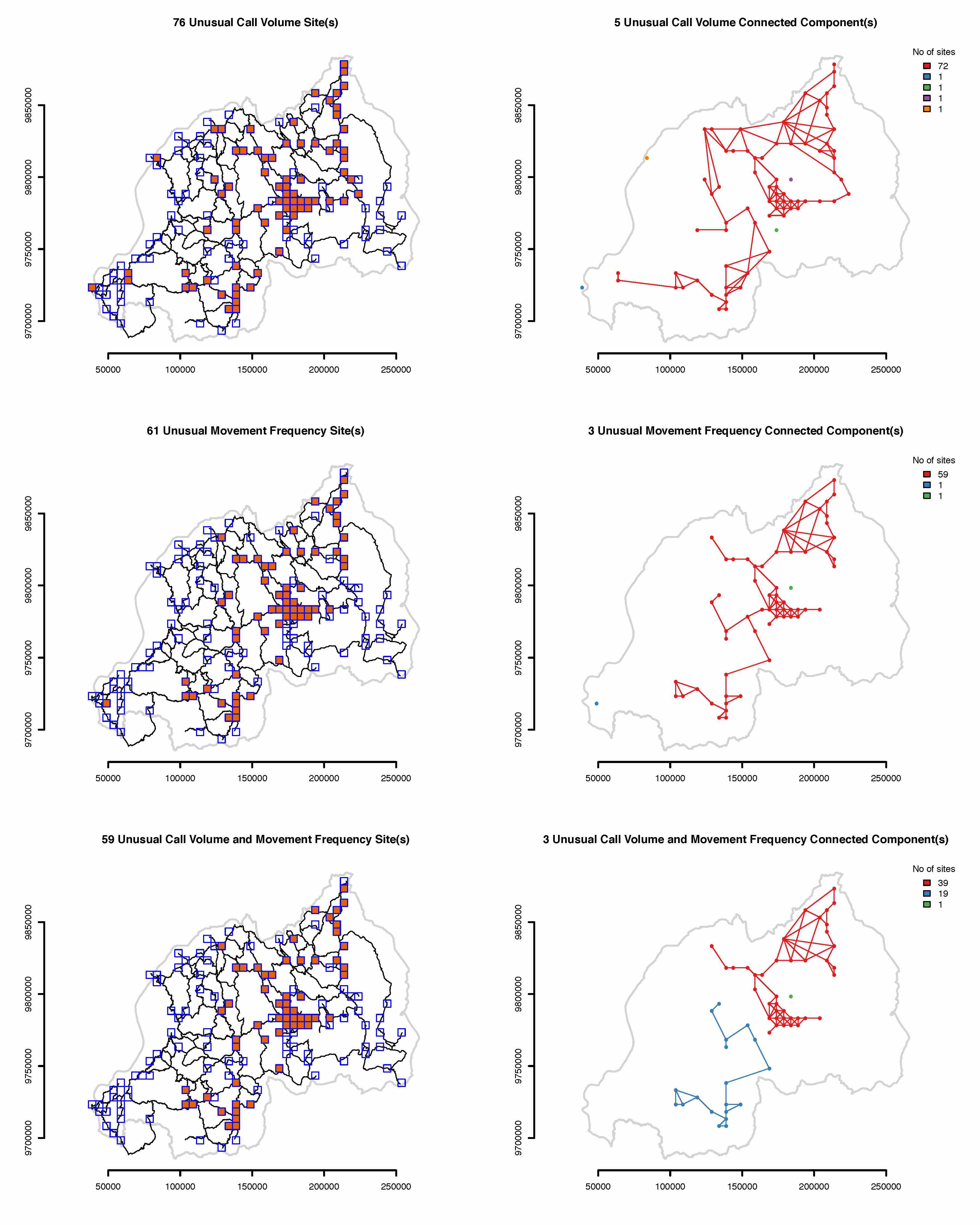}
\caption{{\bf Sites with unusually high behavior on December 24, 2008.}  A number of 59 sites recorded unusually high call volume and movement frequency. A number of 17 additional sites recorded unusually high call volume, while two other sites recorded unusually high movement frequency. Most of the sites in these three groups belong to one or two spatial clusters, but small spatial clusters with only one site are also present.}
\label{fig:UPDecember242008}
\end{center}
\end{figure}

\begin{figure}[htbp!]
\begin{center}
\includegraphics[width=5.5in,angle=0]{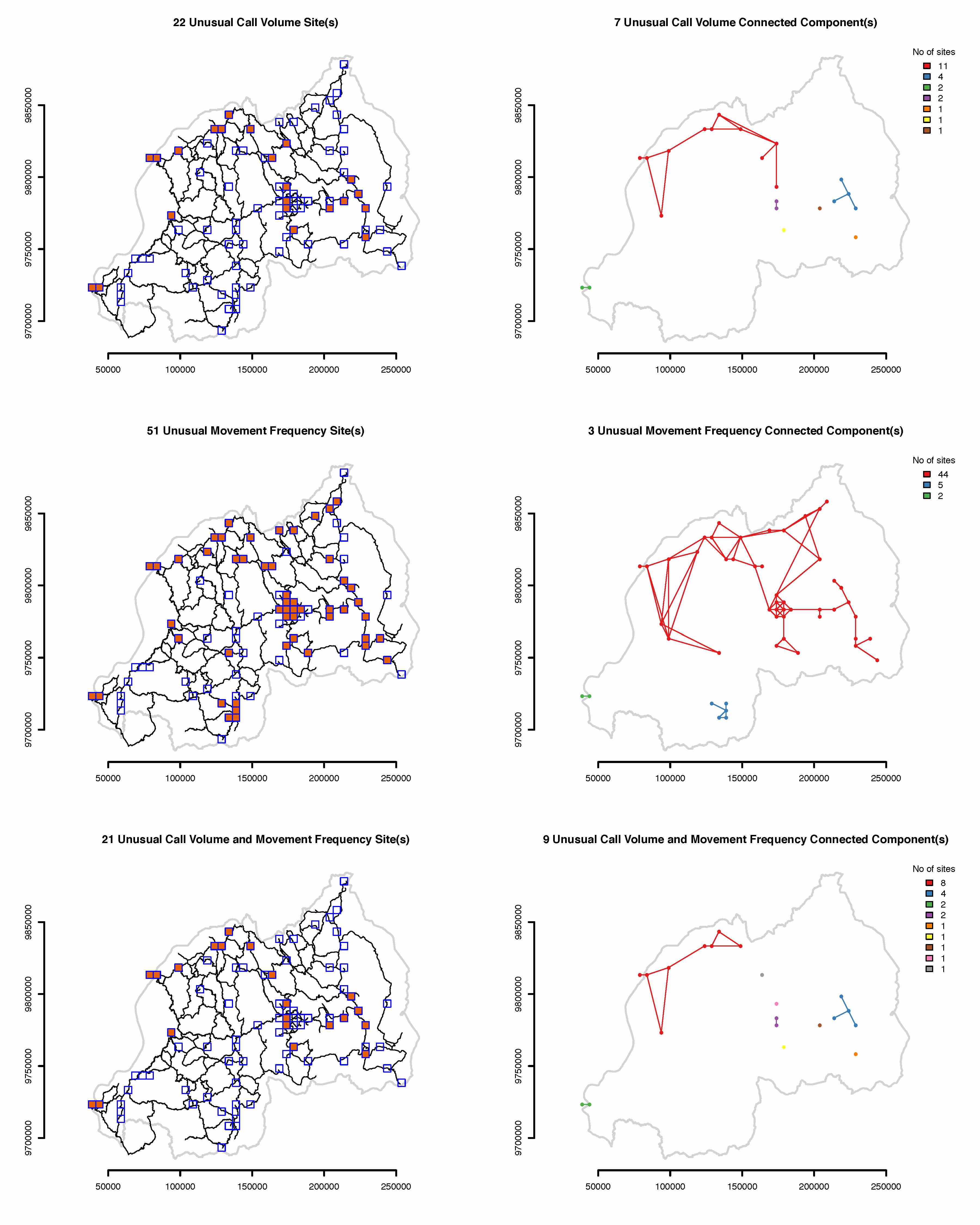}
\caption{{\bf Sites with unusually high behavior on January 1, 2008.}  A number of 21 sites recorded unusually high call volume and movement frequency. One additional site recorded unusually high call volume, while 30 other sites recorded unusually high movement frequency. Most of the sites in these three groups belong to one spatial cluster. Smaller spatial clusters with up to four sites are also present.}
\label{fig:UPJanuary12008}
\end{center}
\end{figure}

\begin{figure}[htbp!]
\begin{center}
\includegraphics[width=5.5in,angle=0]{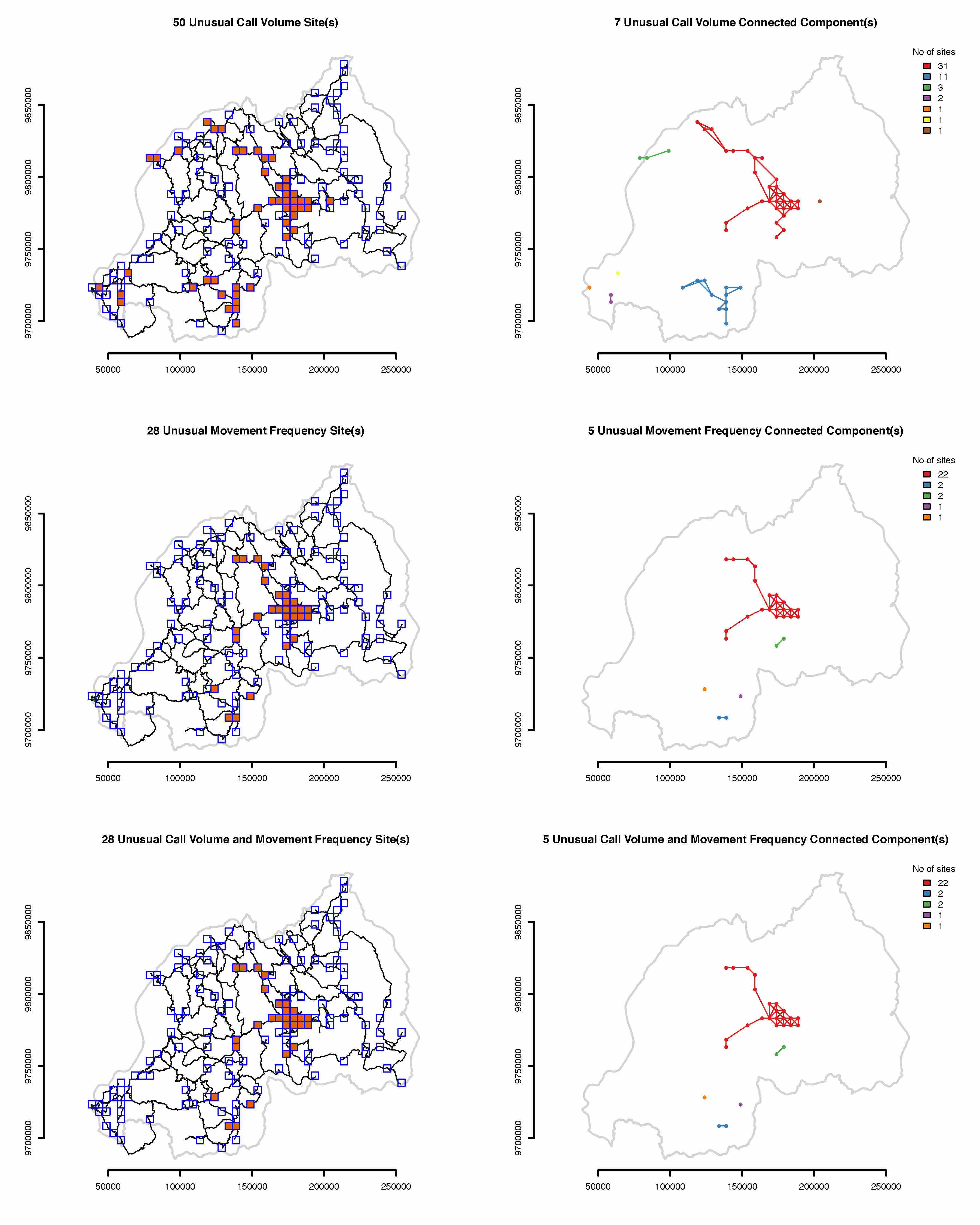}
\caption{{\bf Sites with unusually high behavior on December 31, 2008.}  A number of 28 sites recorded unusually high call volume and movement frequency. A number of 22 additional sites recorded unusually high call volume. Most of the sites in these three groups belong to one or two spatial clusters. Smaller spatial clusters with one or two sites are also present.}
\label{fig:UPDecember312008}
\end{center}
\end{figure}

\begin{figure}[htbp!]
\begin{center}
\includegraphics[width=5.5in,angle=0]{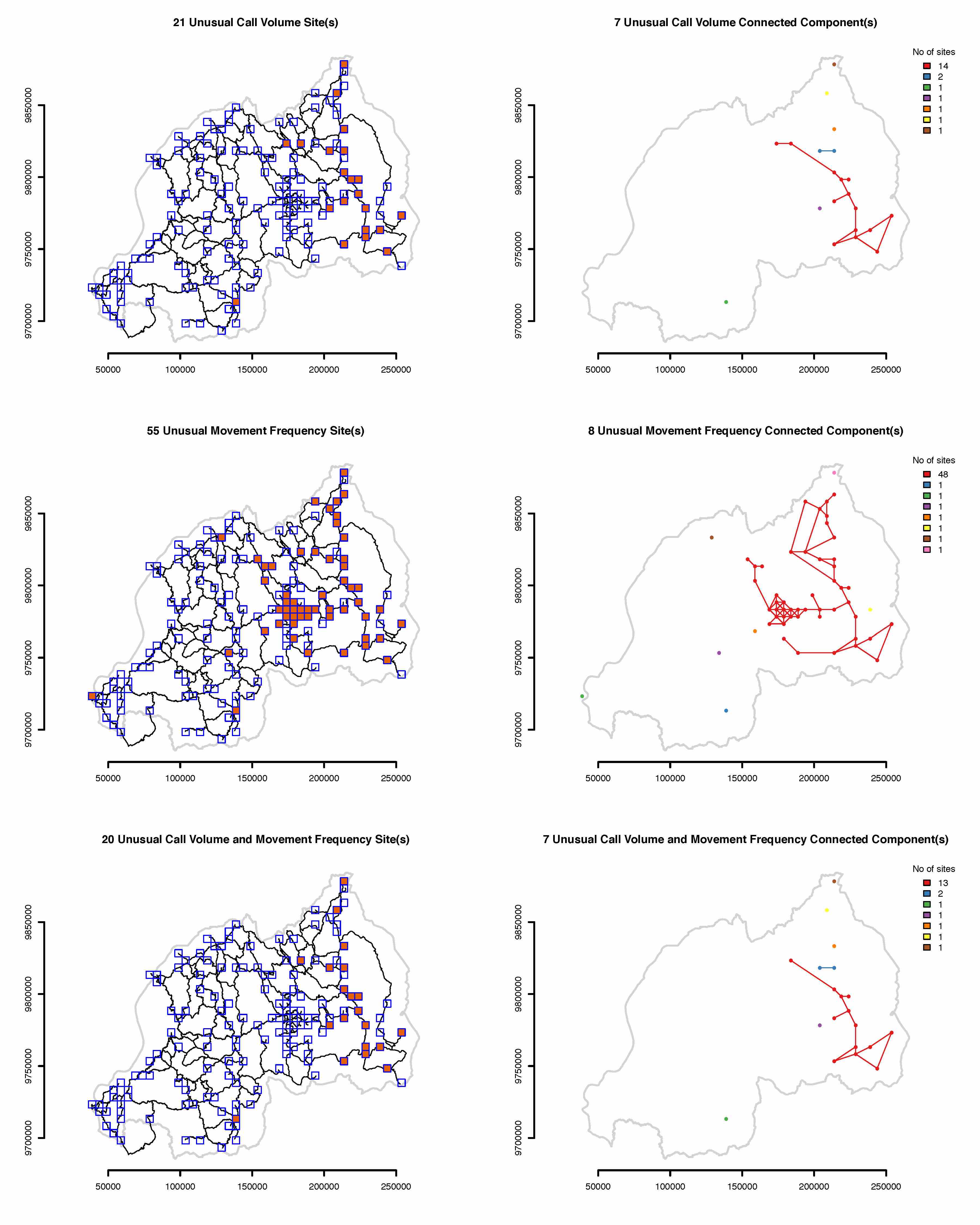}
\caption{{\bf Sites with unusually high behavior on January 1, 2009.}  A number of 20 sites recorded unusually high call volume and movement frequency. One additional site recorded unusually high call volume, while 35 other sites recorded unusually high movement frequency. Most of the sites in these three groups belong to one spatial cluster. Smaller spatial clusters with one or two sites are also present.}
\label{fig:UPJanuary12009}
\end{center}
\end{figure}

\begin{figure}[htbp!]
\begin{center}
\includegraphics[width=5.5in,angle=0]{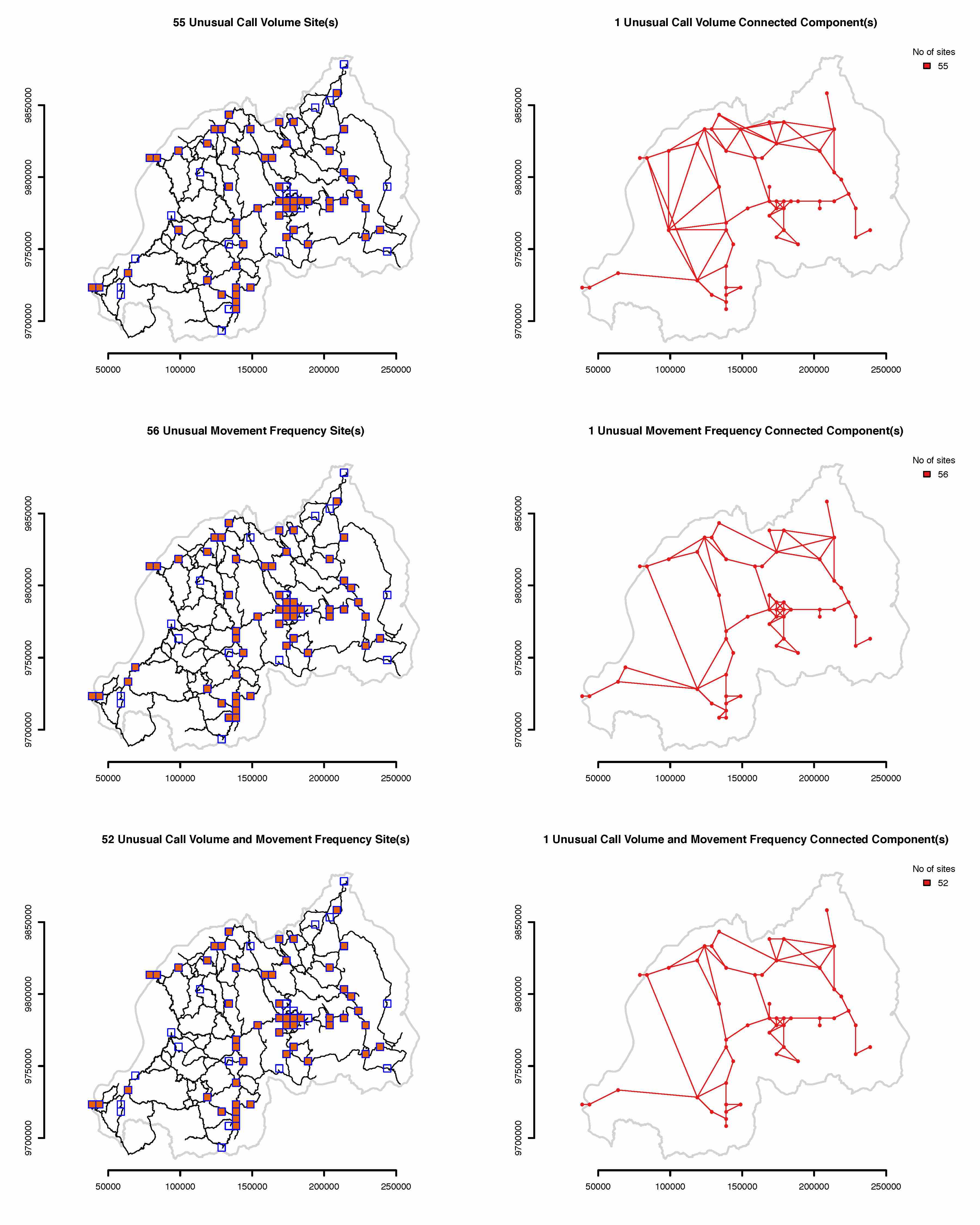}
\caption{{\bf Sites with unusually high behavior on November 9, 2007.}  A number of 52 sites recorded unusually high call volume and movement frequency. Three additional sites recorded unusually high call volume, while one other site recorded unusually high movement frequency. The sites in these three groups belong to one spatial cluster.}
\label{fig:UPNovember92007}
\end{center}
\end{figure}

\begin{figure}[htbp!]
\begin{center}
\includegraphics[width=5.5in,angle=0]{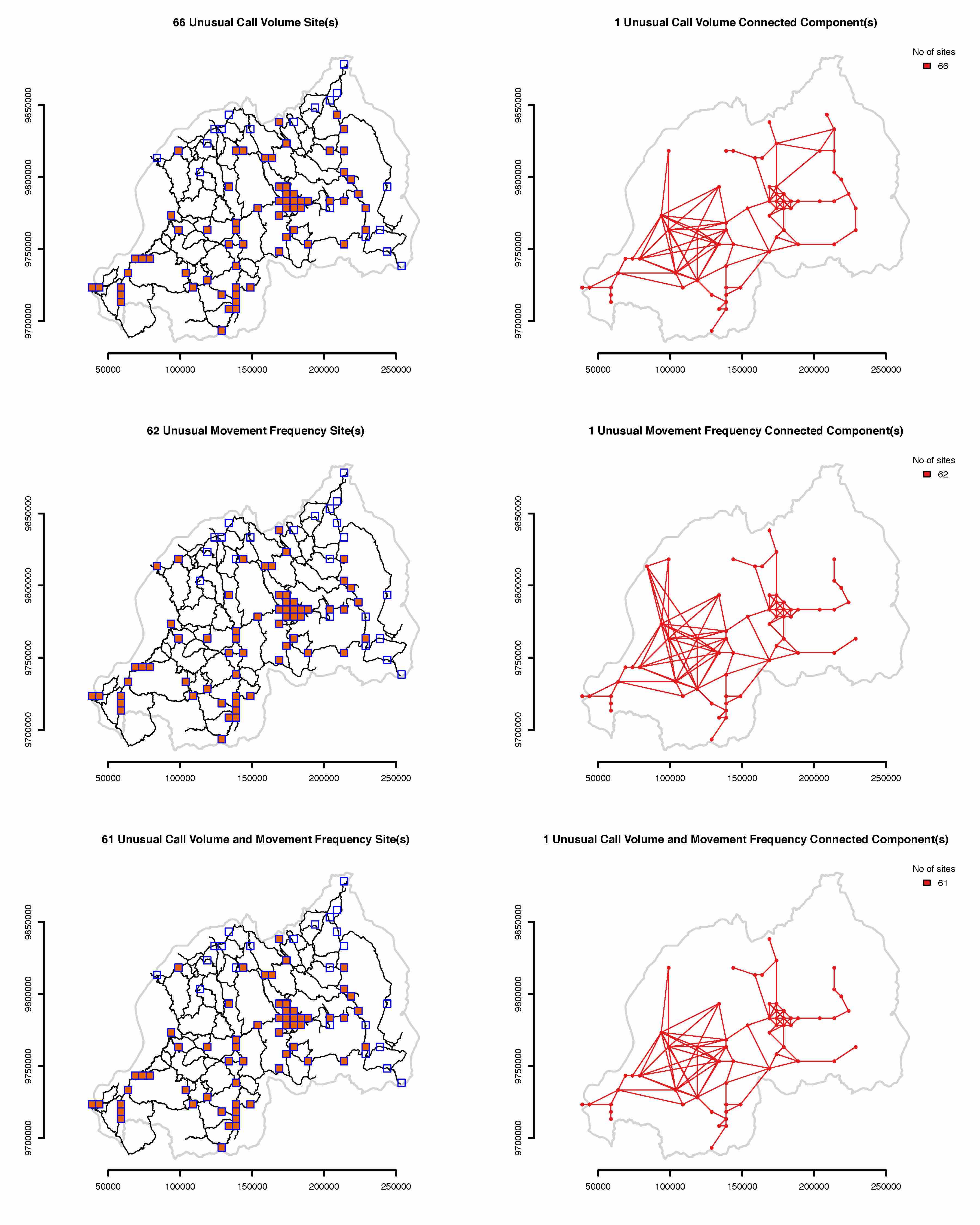}
\caption{{\bf Sites with unusually low behavior on April 24, 2008.}  A number of 61 sites recorded unusually low call volume and movement frequency. Five additional sites recorded unusually low call volume, while one other site recorded unusually low movement frequency. The sites in these three groups belong to one spatial cluster.}
\label{fig:DOWNApril242008}
\end{center}
\end{figure}

\begin{figure}[htbp!]
\begin{center}
\includegraphics[width=5.5in,angle=0]{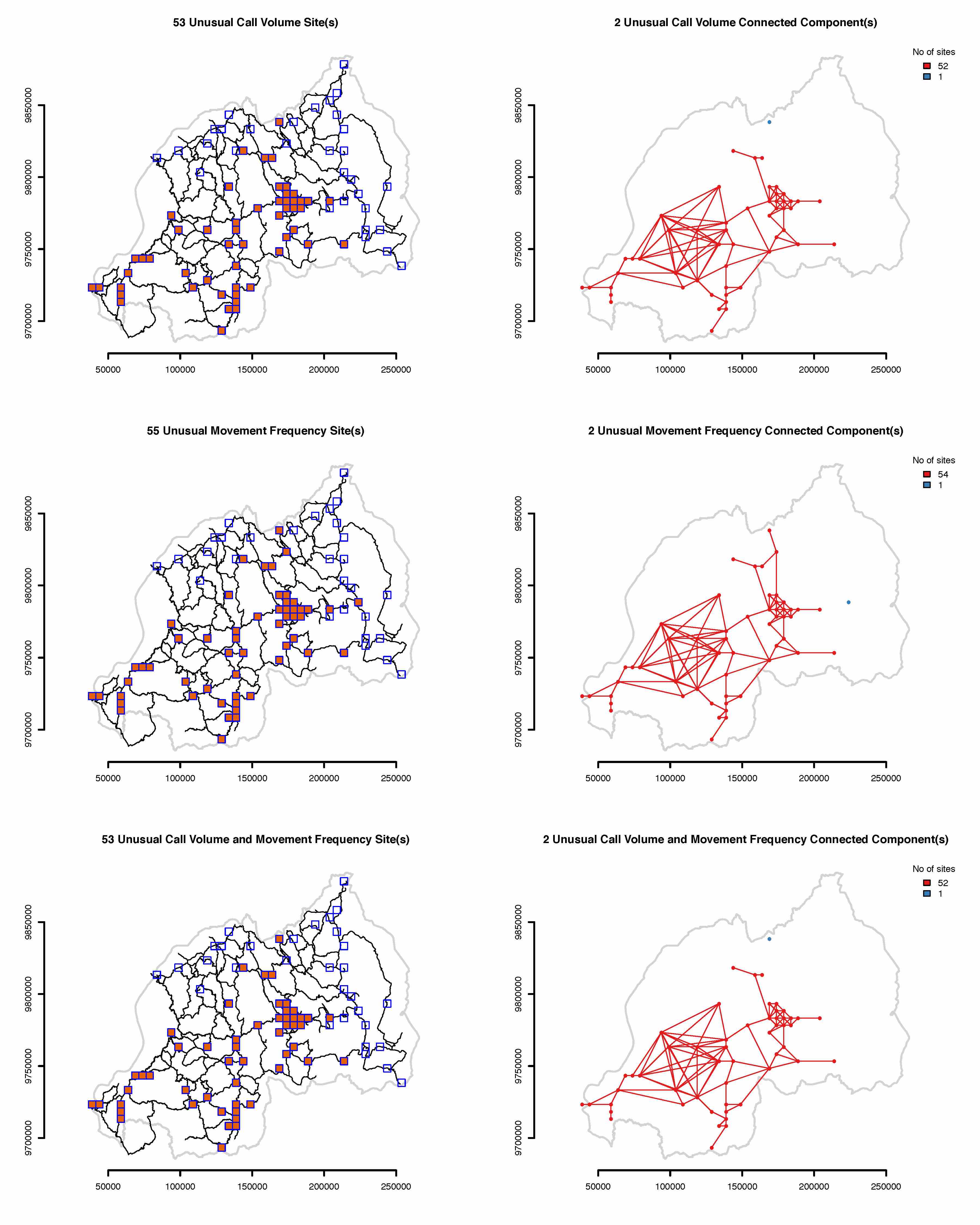}
\caption{{\bf Sites with unusually low behavior on April 25, 2008.}  A number of 53 sites recorded unusually low call volume and movement frequency. Two additional sites recorded unusually low movement frequency. The sites in these two groups belong to one large spatial cluster and to another small spatial cluster with just one site.}
\label{fig:DOWNApril252008}
\end{center}
\end{figure}

\begin{figure}[htbp!]
\begin{center}
\includegraphics[width=5.5in,angle=0]{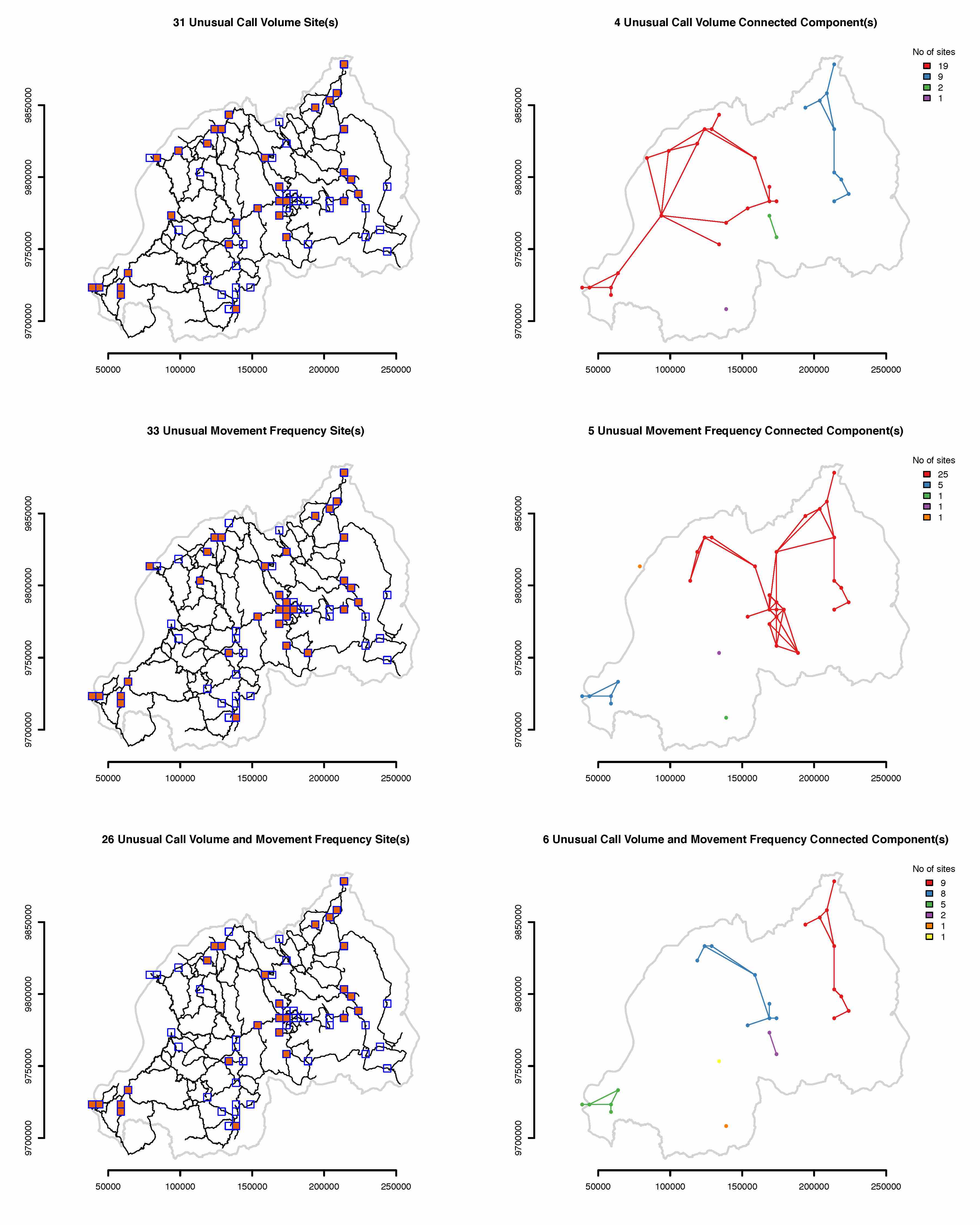}
\caption{{\bf Sites with unusually low behavior on April 7, 2007.}  A number of 26 sites recorded unusually low call volume and movement frequency. Five additional sites recorded unusually low call volume, while 7 other sites recorded unusually low movement frequency. Most of the sites in these three groups belong to two spatial clusters. Smaller spatial clusters are also present.}
\label{fig:DOWNApril72007}
\end{center}
\end{figure}

\begin{figure}[htbp!]
\begin{center}
\includegraphics[width=5.5in,angle=0]{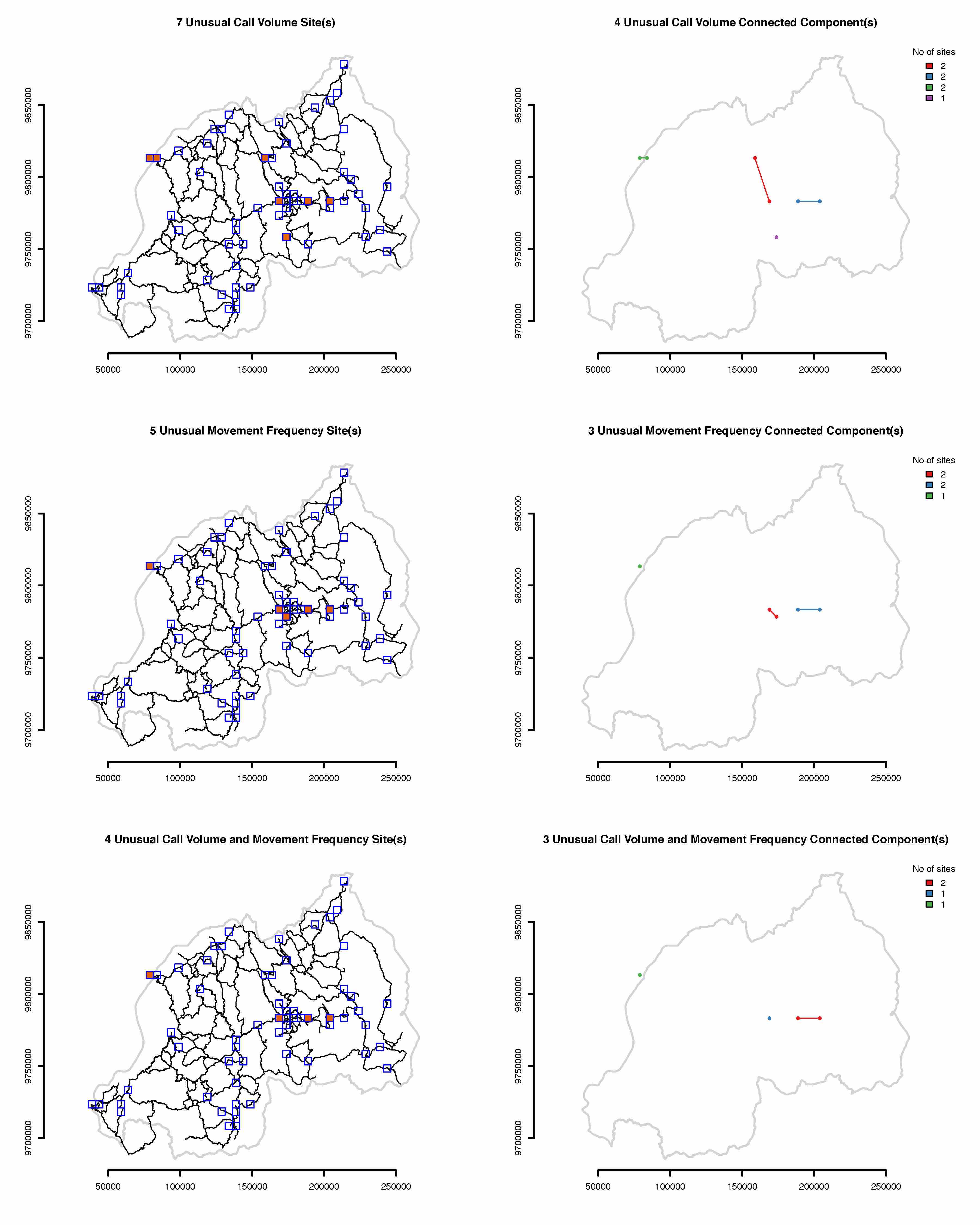}
\caption{{\bf Sites with unusually low behavior on April 8, 2007.}  Four sites recorded unusually low call volume and movement frequency.  Three additional sites recorded unusually low call volume, while one other site recorded unusually low movement frequency. The sites in these three groups belong to spatial clusters of size 1 or 2.}
\label{fig:DOWNApril82007}
\end{center}
\end{figure}

\begin{figure}[htbp!]
\begin{center}
\includegraphics[width=5.5in,angle=0]{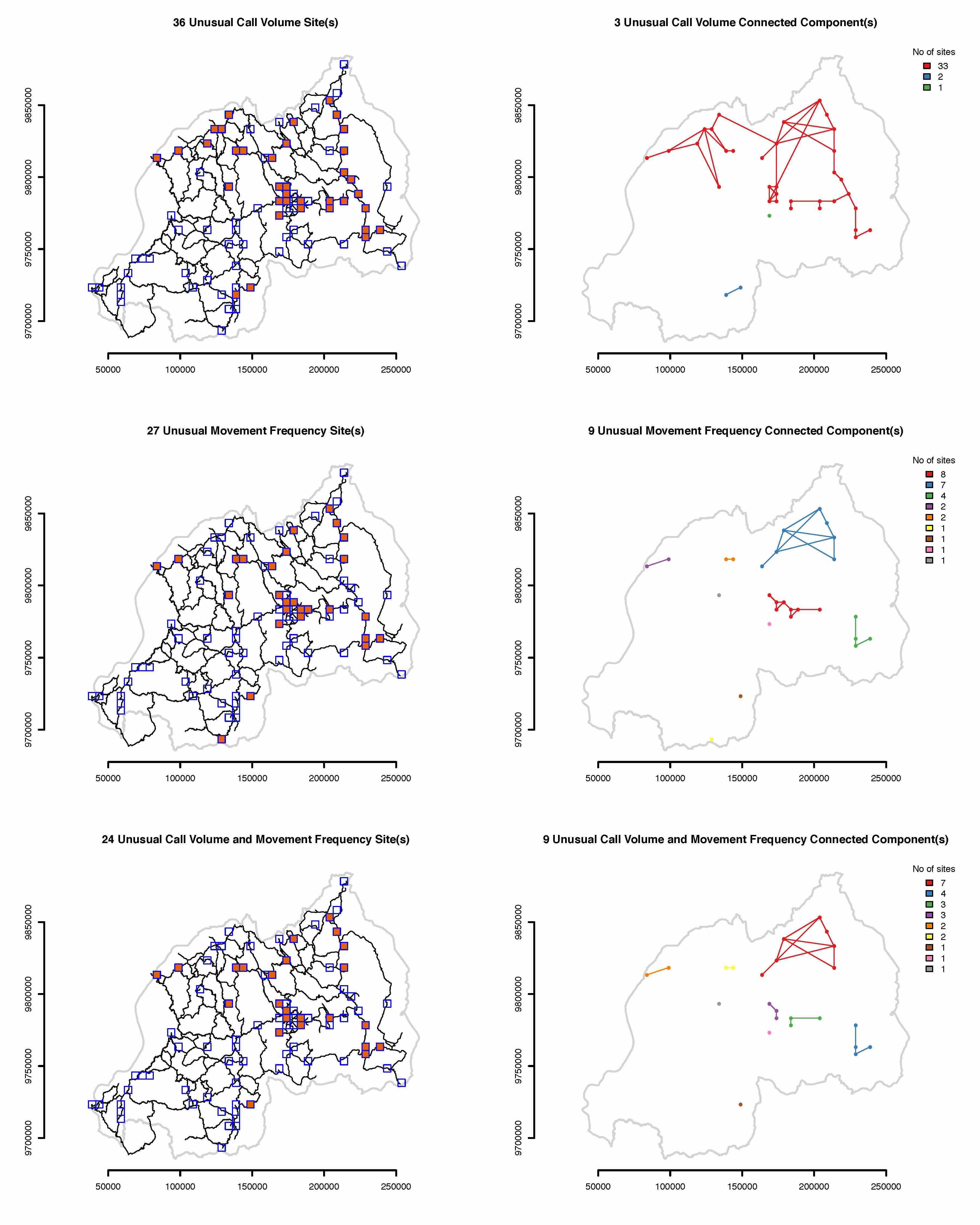}
\caption{{\bf Sites with unusually low behavior on April 7, 2008.}  A number of 24 sites recorded unusually low call volume and movement frequency. Twelve additional sites recorded unusually low call volume, while 3 other sites recorded unusually low movement frequency. Most of the sites in these three groups belong to one spatial cluster. Smaller spatial clusters are also present.}
\label{fig:DOWNApril72008}
\end{center}
\end{figure}

\begin{figure}[htbp!]
\begin{center}
\includegraphics[width=5.5in,angle=0]{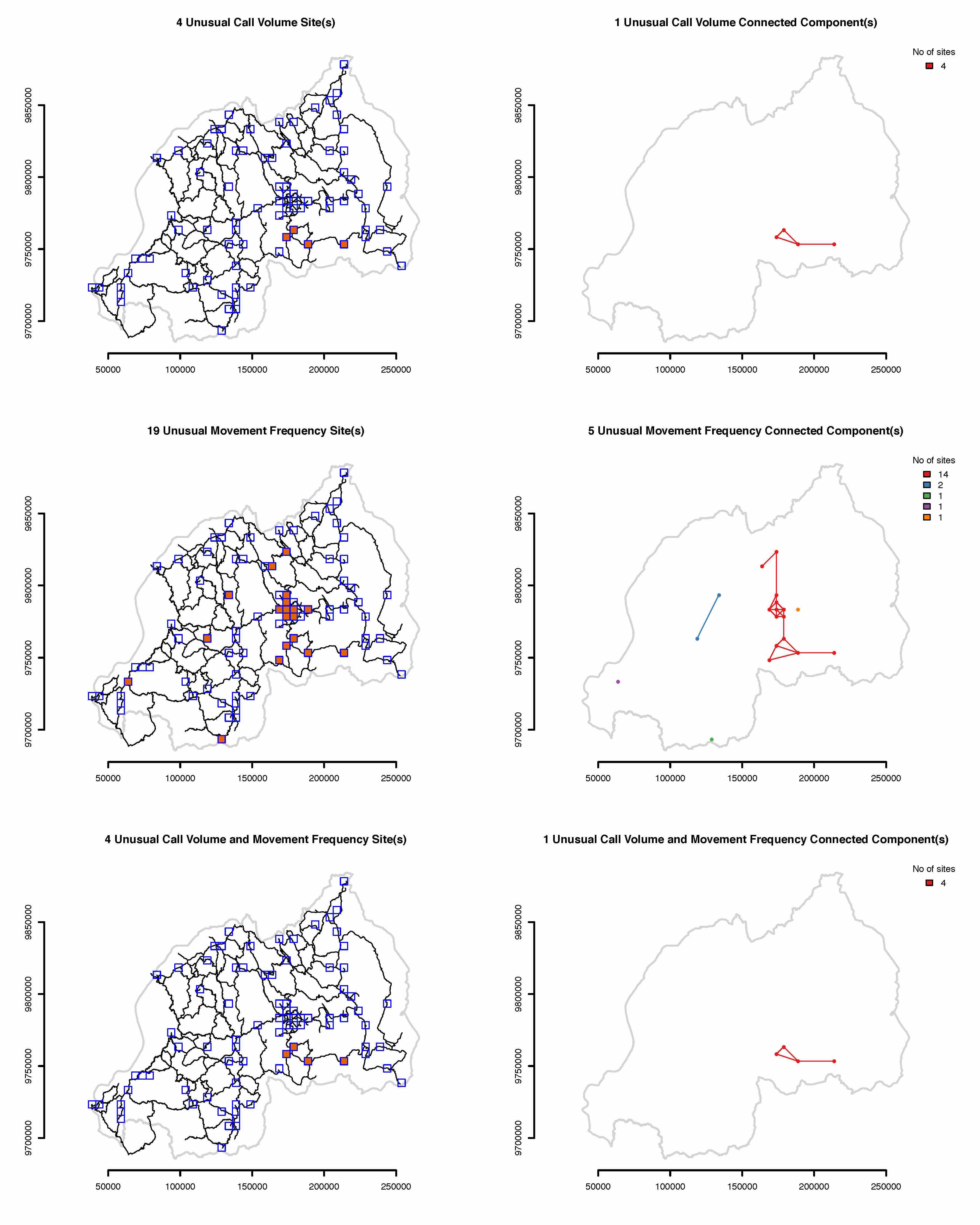}
\caption{{\bf Sites with unusually low behavior on April 8, 2008.} Four sites recorded unusually low call volume and movement frequency. A number of 15 additional sites recorded unusually low movement frequency. Most of the sites in these two groups belong to one spatial cluster. Smaller spatial clusters are also present.}
\label{fig:DOWNApril82008}
\end{center}
\end{figure}




\end{document}